\documentclass[aip,apl,twocolumn,superscriptaddress,10pt]{revtex4-1}

\usepackage{graphicx}

\usepackage{caption}
\usepackage{subcaption}

\usepackage{amsmath}

\usepackage{color}
\definecolor{RED}{rgb}{1.0, 0.0, 0.0}

\usepackage{epstopdf}

\usepackage{multirow,bigstrut}

\usepackage{time,mathptmx,helvet}

\usepackage[
pdfstartview=XYZ,
bookmarks=false,
colorlinks=true,
linkcolor=blue,
urlcolor=blue,
citecolor=blue,
pdftex
]{hyperref}

\makeatletter

\newcommand{\Rmnum}[1]{\expandafter\@slowromancap\romannumeral #1@}
\makeatother

\begin{document}

\title{\large\textcolor{blue}{Direct measurement of non-thermal electron acceleration from magnetically driven reconnection in a laboratory plasma}\vspace{10pt}}

\author{Abraham Chien}
 \email{achien2@pppl.gov}
 \affiliation{Department of Astrophysical Sciences, Princeton University, Princeton, New Jersey 08544 USA}
\author{Lan Gao}
 \affiliation{Princeton Plasma Physics Laboratory, Princeton University, Princeton, New Jersey 08543 USA}
\author{Shu Zhang}
 \affiliation{Department of Astrophysical Sciences, Princeton University, Princeton, New Jersey 08544 USA}
\author{Hantao Ji}
\email{hji@pppl.gov}
 \affiliation{Department of Astrophysical Sciences, Princeton University, Princeton, New Jersey 08544 USA}
 \affiliation{Princeton Plasma Physics Laboratory, Princeton University, Princeton, New Jersey 08543 USA}
\author{Eric G. Blackman}
 \affiliation{Department of Physics and Astronomy, University of Rochester, Rochester, New York 14627 USA}
\author{William Daughton}
 \affiliation{Los Alamos National Laboratory, Los Alamos, New Mexico 87545 USA}
\author{Adam Stanier}
 \affiliation{Los Alamos National Laboratory, Los Alamos, New Mexico 87545 USA}
\author{Ari Le}
 \affiliation{Los Alamos National Laboratory, Los Alamos, New Mexico 87545 USA}
\author{Fan Guo}
 \affiliation{Los Alamos National Laboratory, Los Alamos, New Mexico 87545 USA}
 \author{Russ Follett}
 \affiliation{Laboratory for Laser Energetics, University of Rochester, Rochester, New York 14623 USA}
\author{Hui Chen}
 \affiliation{Lawrence Livermore National Laboratory, Livermore, California 94550 USA}
\author{Gennady Fiksel}
 \affiliation{University of Michigan, Ann Arbor, Michigan 48109 USA}
\author{Gabriel Bleotu}
 \affiliation{ELI-NP, ``Horia Hulubei'' National Institute for Physics and Nuclear Engineering, 30 Reactorului Street, RO-077125, Bucharest-Magurele, Romania}
 \affiliation{University of Bucharest, Faculty of Physics, 077125 Bucharest-Magurele,
Romania}
 \affiliation{LULI-CNRS, CEA, UPMC Univ Paris 06: Sorbonne Universite, Ecole Polytechnique, Institut Polytechnique de Paris, F-91128 Palaiseau Cedex, France}
 \author{Robert C. Cauble}
 \affiliation{Lawrence Livermore National Laboratory, Livermore, California 94550 USA}
 \author{Sophia N. Chen}
 \affiliation{ELI-NP, ``Horia Hulubei'' National Institute for Physics and Nuclear Engineering, 30 Reactorului Street, RO-077125, Bucharest-Magurele, Romania}
 \author{Alice Fazzini}
 \affiliation{LULI-CNRS, CEA, UPMC Univ Paris 06: Sorbonne Universite, Ecole Polytechnique, Institut Polytechnique de Paris, F-91128 Palaiseau Cedex, France}
 \author{Kirk Flippo}
 \affiliation{Los Alamos National Laboratory, Los Alamos, New Mexico 87545 USA}
\author{Omar French}
 \affiliation{University of Maryland, Baltimore County, Baltimore, Maryland 21250 USA}
 \author{Dustin H. Froula}
 \affiliation{Laboratory for Laser Energetics, University of Rochester, Rochester, New York 14623 USA}
\author{Julien Fuchs}
 \affiliation{LULI-CNRS, CEA, UPMC Univ Paris 06: Sorbonne Universite, Ecole Polytechnique, Institut Polytechnique de Paris, F-91128 Palaiseau Cedex, France}
 \author{Shinsuke Fujioka}
 \affiliation{Institute of Laser Engineering, Osaka University, Osaka, 565-0871, Japan}
\author{Kenneth Hill}
 \affiliation{Princeton Plasma Physics Laboratory, Princeton University, Princeton, New Jersey 08543 USA}
 \author{Sallee Klein}
 \affiliation{University of Michigan, Ann Arbor, Michigan 48109 USA}
 \author{Carolyn Kuranz}
 \affiliation{University of Michigan, Ann Arbor, Michigan 48109 USA}
 \author{Philip Nilson}
 \affiliation{Laboratory for Laser Energetics, University of Rochester, Rochester, New York 14623 USA}
\author{Alexander Rasmus}
 \affiliation{Los Alamos National Laboratory, Los Alamos, New Mexico 87545 USA}
\author{Ryunosuke Takizawa}
 \affiliation{Institute of Laser Engineering, Osaka University, Osaka, 565-0871, Japan}

\date{\today}

\begin{abstract}

Magnetic reconnection is a ubiquitous astrophysical process that rapidly converts magnetic energy into some combination of plasma flow energy, thermal energy, and non-thermal energetic particles, including energetic electrons~\cite{yamada10,ji22}.  Various reconnection acceleration mechanisms~\cite{blandford17,guo20} in different low-$\beta$ (plasma-to-magnetic pressure ratio) and collisionless environments~\cite{masuda94,oieroset02,krucker10,abdo11,tavani11,chen20} have been proposed theoretically and studied numerically~\cite{drenkhahn02,sironi14,werner16,li17,guo20,dahlin20}, including first- and second-order Fermi acceleration~\cite{drake06}, betatron acceleration~\cite{hoshino01}, parallel electric field acceleration along magnetic fields~\cite{egedal13}, and direct acceleration by the reconnection electric field~\cite{zenitani01}. However, none of them have been heretofore confirmed experimentally, as the direct observation of non-thermal particle acceleration in laboratory experiments has been difficult due to short Debye lengths for \textit{in-situ} measurements and short mean free paths for \textit{ex-situ} measurements. Here we report the direct measurement of accelerated non-thermal electrons from low-$\beta$ magnetically driven reconnection in experiments using a laser-powered capacitor coil platform.  We use kiloJoule lasers to drive parallel currents to reconnect MegaGauss-level magnetic fields in a quasi-axisymmetric geometry~\cite{Gao2016,chien2019,chien21}. The angular dependence of the measured electron energy spectrum and the resulting accelerated energies, supported by particle-in-cell simulations, indicate that the mechanism of direct electric field acceleration by the out-of-plane reconnection electric field~\cite{zenitani01,uzdensky11,cerutti13} is at work. Scaled energies using this mechanism show direct relevance to astrophysical observations. Our results therefore validate one of the proposed acceleration mechanisms by reconnection, and establish a new approach to study reconnection particle acceleration with laboratory experiments in relevant regimes.
\end{abstract}

\maketitle

Magnetic reconnection, the process by which magnetic field topology in a plasma is reconfigured, rapidly converts magnetic energy into some combination of bulk flow, thermal, and accelerated particles.  The latter is a prominent feature of presumed reconnection regions in nature, and as such, reconnection can be thought of as an efficient particle accelerator in low-$\beta$ ($\lesssim 1$), collisionless plasmas where abundant magnetic free energy per particle is available. Electron acceleration up to $\sim300~\textrm{keV}$, for example, has been observed in Earth's magnetotail~\cite{oieroset02} and the measured spectra in X-ray, extreme ultraviolet, and microwave wavelengths from solar flares include a non-thermal power law component, indicating a large supra-thermal electron population~\cite{masuda94,krucker10,chen20}. Reconnection has been suggested as the underlying source of these non-thermal electrons. Gamma-ray flares from the Crab Nebula are another example, exhibiting particle acceleration up to $10^{15}~\textrm{eV}$, which cannot be explained by shock acceleration mechanisms~\cite{tavani11,abdo11,kroon16}. 

The efficient acceleration of charged particles by magnetic reconnection has been studied theoretically and numerically\cite{sironi14,dahlin2014,werner16,dahlin2016,totorica2016,li17}, and various acceleration mechanisms, such as parallel electric field acceleration and Fermi acceleration, have been proposed~\cite{zenitani01,hoshino01,drake06,egedal13}. However, thus far, no direct measurements of non-thermal particle acceleration due to low-$\beta$ reconnection have been made in laboratory experiments to confirm or contradict these mechanisms. Short Debye lengths and mean free paths have limited most \textit{in-situ} and \textit{ex-situ} detection of the predicted energetic electrons, respectively, while indirect measurements of energetic electrons are necessarily limited by specific models assumed for radiation and acceleration mechanisms~\cite{savrukhin01,klimanov07,dubois2017}.

\begin{figure*}[htb]
  \begin{center}
    \includegraphics[width=0.8\textwidth]{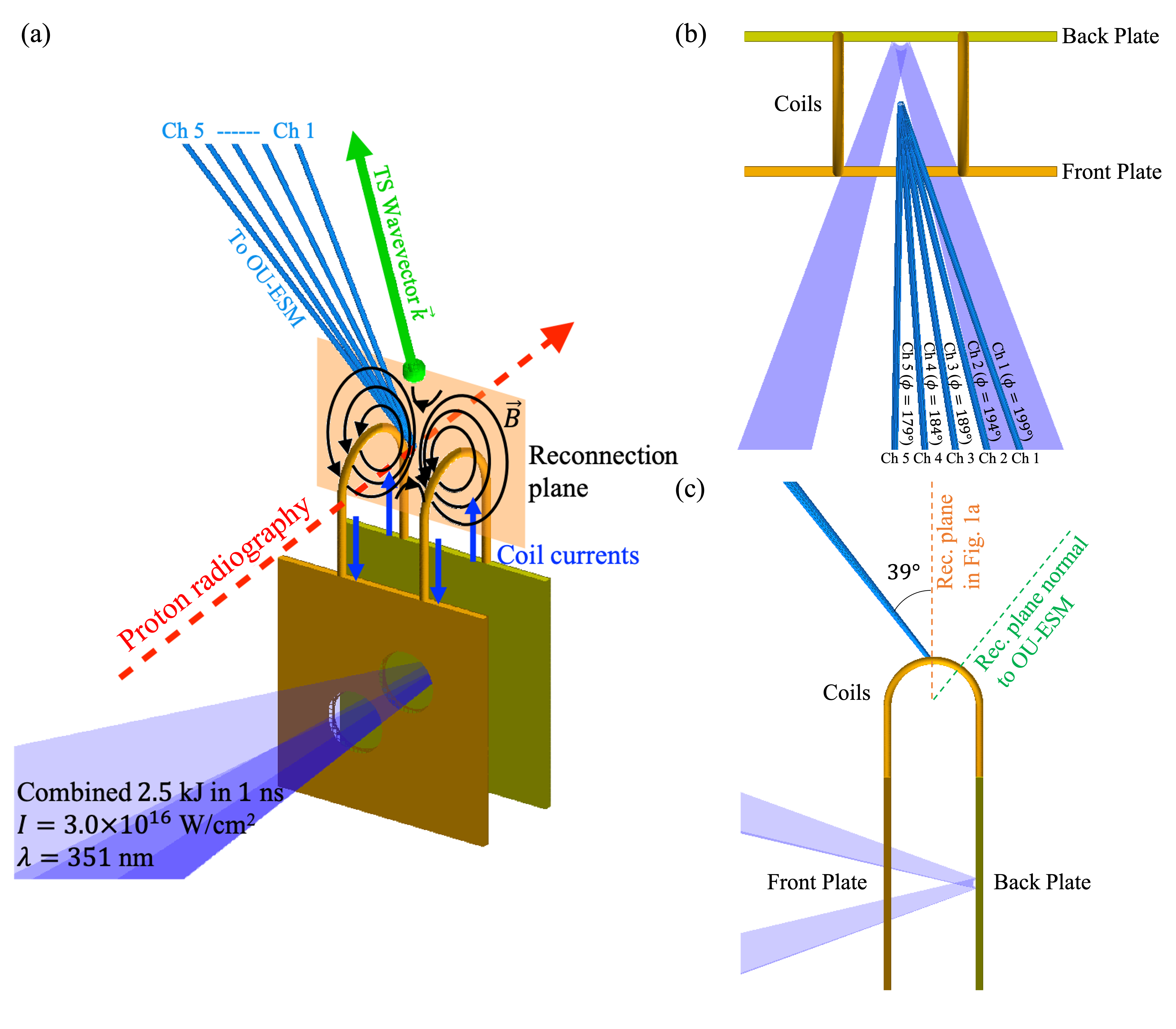}
    \caption{\small \textbf{Experimental setup of low-$\beta$, magnetically driven reconnection using laser-powered capacitor coils.} \textbf{a,} The capacitor-coil target is driven by two long-pulse lasers, passing through the front holes and irradiating the back plate. An electrostatic potential is created between the (capacitor) plates, and large currents (blue arrows) are generated in the parallel U-shaped coils. The resulting magnetic fields form a reconnection structure between the coils. Major diagnostics are target-normal sheath acceleration (TNSA) proton radiography and the Osaka University electron spectrometer (OU-ESM). The OU-ESM is positioned 37.5 cm away from the main interaction, at an angle 39$^\circ$ from the vertical. 5 independent channels are situated with 5$^\circ$ between each channel, allowing a measurement of the angular spread of electrons in the azimuthal direction. Thomson scattering measurements were taken in a similar experiment to diagnose plasma parameters: the green ball shows the probing volume, and the Thomson scattering wavevector $\vec{k}$ is also shown. \textbf{b,} A top-down view of the main target is shown, along with the OU-ESM channel orientation in the azimuthal direction. \textbf{c,} A side-on view of the main target shows the relative polar orientation of the OU-ESM channels. The orange vertical dashed line represents the reconnection plane shown in \textbf{a}, and the green dashed line represents the reconnection plane that is normal to the OU-ESM line of sight.}
    \label{fig:1}
  \end{center}
\end{figure*}

High-energy-density (HED) plasmas~\cite{nilson06,willingale10,zhong10,fiksel14,raymond2018,Gao2016,chien2019,chien21,dong12,Zhong_2016} have recently emerged as novel platforms to study magnetic reconnection. In particular, direct measurements of charged particle spectra are possible due to a large electron mean free path relative to the detector distance. Importantly, low-$\beta$, collisionless, magnetically driven reconnection is achievable using laser-powered capacitor coils~\cite{Gao2016,chien2019,chien21}, allowing relevant conditions to astrophysical environments. Here, using this experimental reconnection platform, we directly detect non-thermal electron acceleration from reconnection, and combined with particle-in-cell simulations, infer a primary acceleration mechanism of direct electric field acceleration by the reconnection electric field.

\begin{figure*}[tb]
  \begin{center}
    \includegraphics[width=1.0\textwidth]{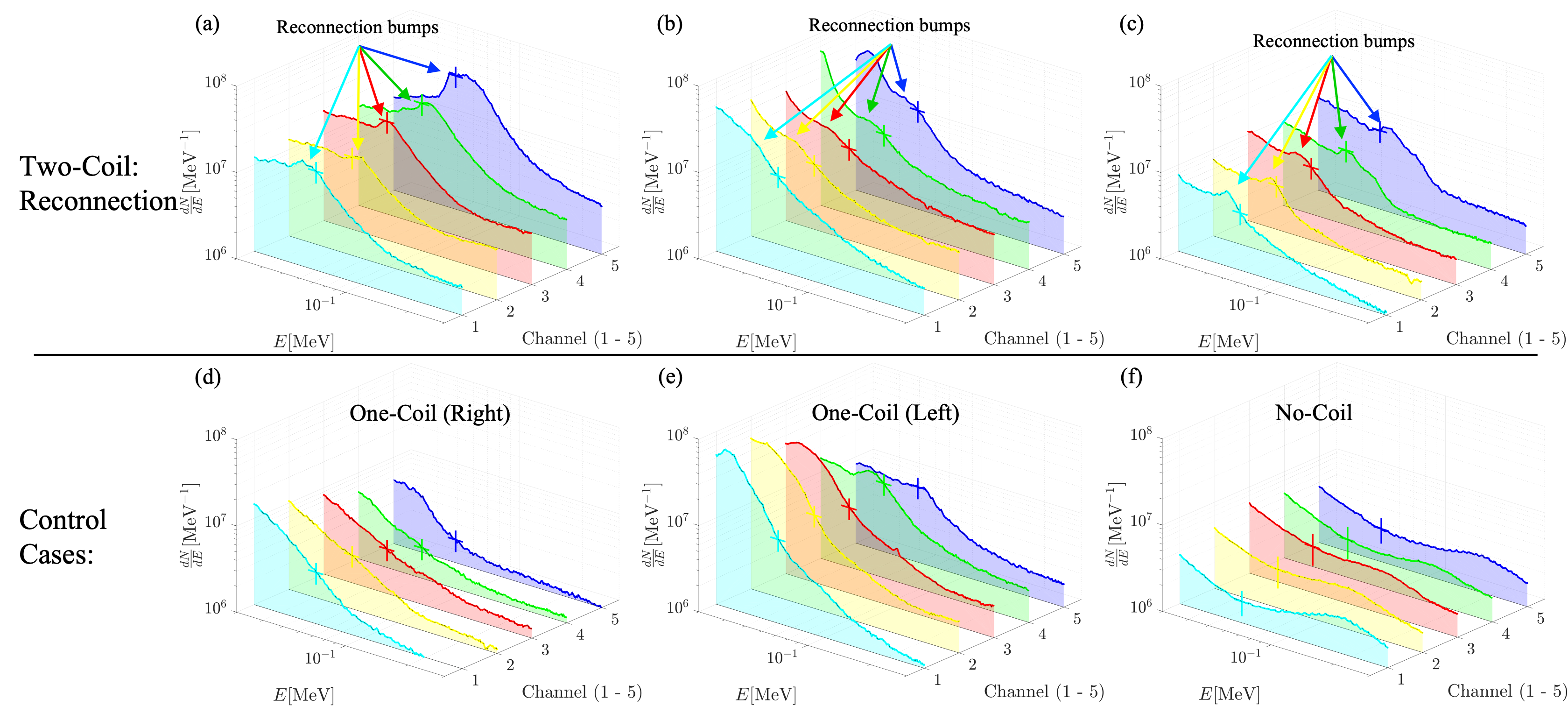}
    \caption{\small \textbf{Experimental particle spectra exhibit conspicuous evidence for non-thermal acceleration bumps in reconnection cases, that are absent in non-reconnection control cases.} Particle spectra from the OU-ESM are presented for 6 experimental shots: 3 two-coil reconnection cases (\textbf{a, b, c}), 2 one-coil control cases (\textbf{d, e}), and 1 no-coil control case (\textbf{f}). 5 colors represent the 5 channels spread in azimuthal angle (Fig. 1b). In all plots, a cross represents characteristic horizontal and vertical error bars at $E=60~\textrm{keV}$. Despite shot-to-shot variations in signal level, in the reconnection cases, spectral bumps are observed in the $40–70~\textrm{keV}$ range. These bumps are significant and observed to exceed the experimental signal error. They are strongest in Channel 5, representing a near-face-on view of the target, and decrease with larger azimuthal angle, with the weakest bumps in Channel 1, representing $19^{\circ}$ off normal. Such a trend is noticeably absent in the control cases, where overall weaker signal levels are observed. A feature appearing to be a spectral bump is observed in \textbf{e}, but it is, in fact, a deficiency in the low-energy range: due to the coil position on the left, low-energy electrons are preferentially deflected toward large $\theta-z$ pitch angle, resulting in an absence in Channels 4 and 5. The no-coil control case (\textbf{f}) exhibits significantly lower signal level than the one-coil control cases and reconnection cases, implying the magnetic field from the coils is deflecting hot electrons toward the detector.}
    \label{fig:2}
  \end{center}
\end{figure*}

Our experiments using laser-powered capacitor coils were performed at the OMEGA EP facility at the Laboratory for Laser Energetics (LLE). The experimental setup, with diagnostic locations, is shown in Fig.~\ref{fig:1}. The capacitor-coil target is driven with two laser pulses, each delivering 1.25-kiloJoule of laser energy in a 1-ns
square temporal profile at a wavelength of 351 nm. The corresponding on-target laser intensity is $\sim3\times10^{16}~\textrm{W}/\textrm{cm}^{2}$. Due to the laser interaction, strong currents are driven in the coils. In targets with two parallel coils, a magnetic reconnection field geometry is created between the coils, and in targets with one coil, a simple magnetic field around a wire is produced, representing a non-reconnection control case. Further information on capacitor coil target operation and design are provided in the Methods section.

The coil current profile can be approximated by a linear rise during the laser pulse ($0<t<t_{\textrm{rise}}$), followed by an exponential decay after laser turn-off. Target sheath normal acceleration (TNSA)~\cite{wilks2001} proton radiography measurements indicated a maximum coil current at $t_{\textrm{rise}}=1~\textrm{ns}$ of $57~\textrm{kA}$, corresponding to a magnetic field at the center of the coils of $110~\textrm{T}$ and an upstream reconnection magnetic field strength of $50.7~\textrm{T}$, with a subsequent exponential decay time of $t_{\textrm{decay}}=8.6~\textrm{ns}$~\cite{chien2019}. During the current rise, the magnetic field strengthens, driving ``push''-phase reconnection, where field lines are pushed into the reconnection region, and during current decay, ``pull'' reconnection occurs, where field lines are pulled out of the reconnection region~\cite{YamadaPOP1997}. Due to the short timescale of the push phase relative to the pull phase, reconnection is driven more strongly during the push phase, and the push phase is the dominant source of particle acceleration.

We used Thomson scattering to diagnose the reconnecting copper plasma in a similar experiment on the OMEGA laser, and found electron density $n_{e}\simeq 3\times10^{18}~\textrm{cm}^{-3}$, ion density $n_{i}\simeq 1.7\times10^{17}~\textrm{cm}^{-3}$, and electron and ion temperatures $T_{e}\simeq T_{i} \simeq 400~\textrm{eV}$. Due to the large $Z=18$, the ion plasma pressure is negligible compared to the electron plasma pressure, and the ratio of plasma pressure to magnetic pressure $\beta \simeq 0.05$. The experiments are therefore firmly in the low-$\beta$ regime, most pertinent for particle acceleration in astrophysical conditions. The Lundquist number is $10^{3}-10^{4}$, representing collisionless reconnection. The reconnection system size is defined by the inter-coil distance of $L=600~\mu\textrm{m}$, and when normalized by the ion skin depth $d_i$, the normalized system size $L/d_{i} \simeq 1.4$. Due to the small system size, the reconnection is deeply in the electron-only regime~\cite{phan18}, where ions are decoupled.

A time-integrated electron spectrometer -- the Osaka University electron spectrometer (OU-ESM) -- was used to measure the electron energy spectra. It is located $37.5~\textrm{cm}$ away from the coils, at a polar angle of $39^{\circ}$ and scans an azimuthal range of $179^{\circ}-199^{\circ}$ with five equally-spaced detection channels. The OU-ESM channel orientation is shown in Fig.~\ref{fig:1}(b,c). Further details regarding the OU-ESM are given in the Methods section. Fig.~\ref{fig:2} shows the experimentally measured OU-ESM data for three double-coil reconnection shots as well as two single-coil and one no-coil control shots. While neither one-coil nor no-coil shots represent perfect control cases, the combination of these configurations allows for better isolation of the reconnection signal.

\begin{figure*}[htb]
  \begin{center}
    \includegraphics[width=0.9\textwidth]{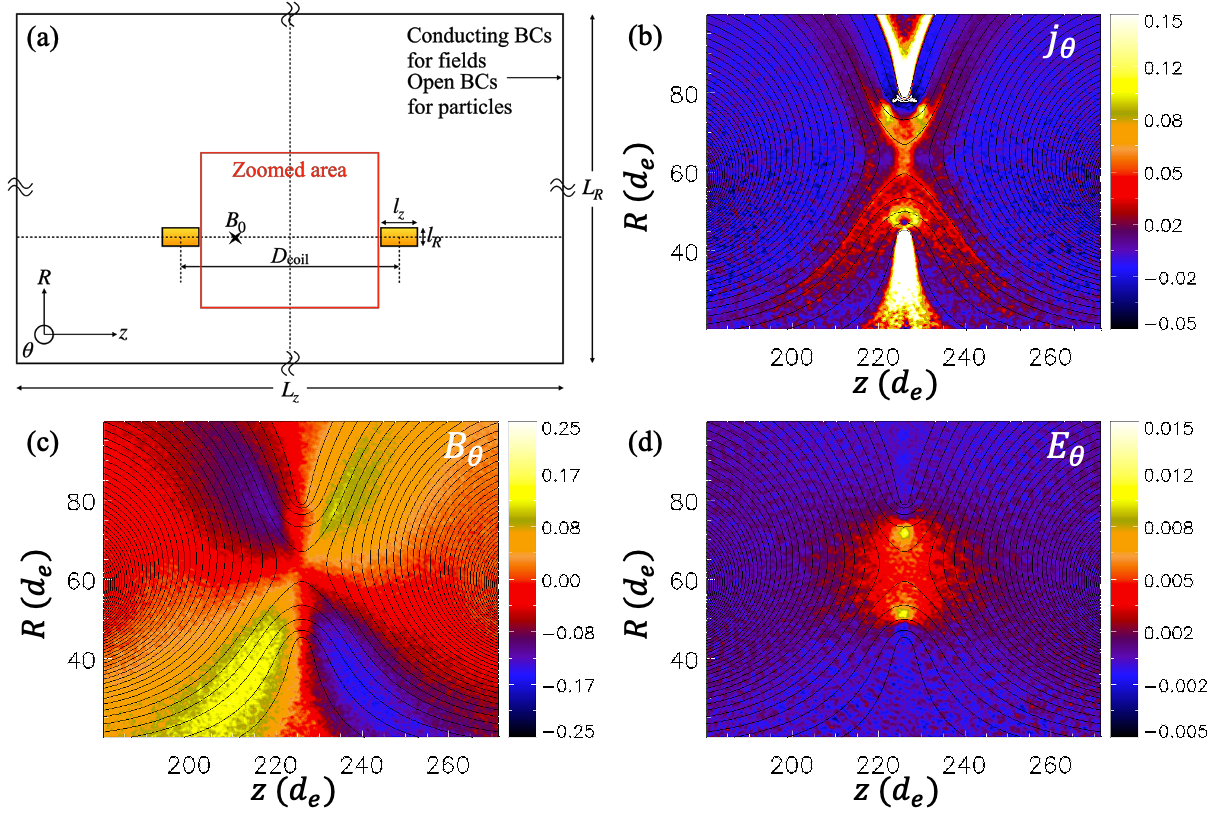}
    \caption{\small \textbf{Particle-in-cell simulation setup of capacitor-coil reconnection demonstrates push reconnection and quadrupole out-of-plane magnetic fields.} \textbf{a,} Schematic of the 2-D cylindrical simulation box used in VPIC modeling, with $z$ the axis of symmetry and $R$ the radial direction. The azimuthal angle $\theta$ is directed out-of-the-page. Two rectangular coils are situated at $R=R_{coil}$ with center-to-center separation of $D_{coil}$. The reference magnetic field $B_{0}$ is measured upstream from the coils. Conducting boundary conditions for fields and open boundary conditions for particles are stipulated. Current is injected in the coils with time, replicating the experimental current profile. Time evolution of fields are evaluated relative to the current rise time $t_{\textrm{rise}}$, to keep the magnetic field drive consistent. All measurements are taken in the “zoomed” box area, marked in red. The outer box boundary is not to scale. \textbf{b, c, d,} Profiles of out-of-plane electric current density $j_\theta$, out-of-plane magnetic field $B_\theta$, and reconnection electric field $E_\theta$ in the red rectangular zoomed area are shown at $t=1.55~t_{rise}$, overlaid with magnetic field lines. $B_\theta$ shows a characteristic quadrupole field structure from decoupled electron and ion flows. A noticeable reconnection electric field is observed around the current sheet near the magnetic null, with the orientation indicating push reconnection. $j_{\theta}$, $B_{\theta}$, and $E_{\theta}$ are shown in the respective normalized units: $j_{0}=en_{e0}c$, $B_{0}=m_{e}\omega_{pe}/e$, and $E_{0}=cB_{0}=m_{e}\omega_{pe}c/e$.}
    \label{fig:3}
  \end{center}
\end{figure*}

Small differences in the laser energy profile and target properties among shots causes variations in otherwise nominally identical cases as seen in Fig.~\ref{fig:2}.  However, focusing on the angular dependence across the channels for each shot reveals a key feature in the electron spectra: non-thermal ``bumps'' in the reconnection cases that do not appear in the control cases. The bumps span the $50-70~\textrm{keV}$ range, and they are most pronounced at the near-normal Channel 5 ($\phi=179^{\circ}$) and weaken with increasing angle from normal. In contrast, the one-coil control cases do not exhibit consistent spectral bumps, and generally exhibit lower signal level. One exception is Figure~\ref{fig:2}e, which represents a one-coil shot with the coil on the left side (as viewed from the front of the target). Due to the coil magnetic field, low-energy electrons are deflected toward higher $\phi$, resulting in an electron deficiency in Channel 5 and to a lesser extent, Channel 4. The no-coil control case exhibits an even lower signal level than the one-coil case: this is due to the lack of a magnetic field to deflect electrons toward the detector. The background ``thermal'' signal does not represent the $T_{e}=400~\textrm{eV}$ plasma: it is the quasi-Maxwellian suprathermal distribution with a hot ``temperature'' of $T_{e,h}\simeq 40-50~\textrm{keV}$, created by laser-plasma instabilities (LPI), such as stimulated Raman scattering (SRS) and two-plasmon decay (TPD)~\cite{drake_1984,figueroa_1984,ebrahim_1980}. 

These spectral bumps demonstrate non-thermal electron acceleration, and the detection angle dependence of the bump sizes suggests a directional anisotropy in the accelerated electron population. The strongest non-thermal population is seen in the direction out of the reconnection plane, anti-parallel to the reconnection electric field, indicating its responsibility for the direct acceleration~\cite{zenitani01,uzdensky11,cerutti13}. 

\begin{figure*}[htb]
  \begin{center}
    \includegraphics[width=1.0\textwidth]{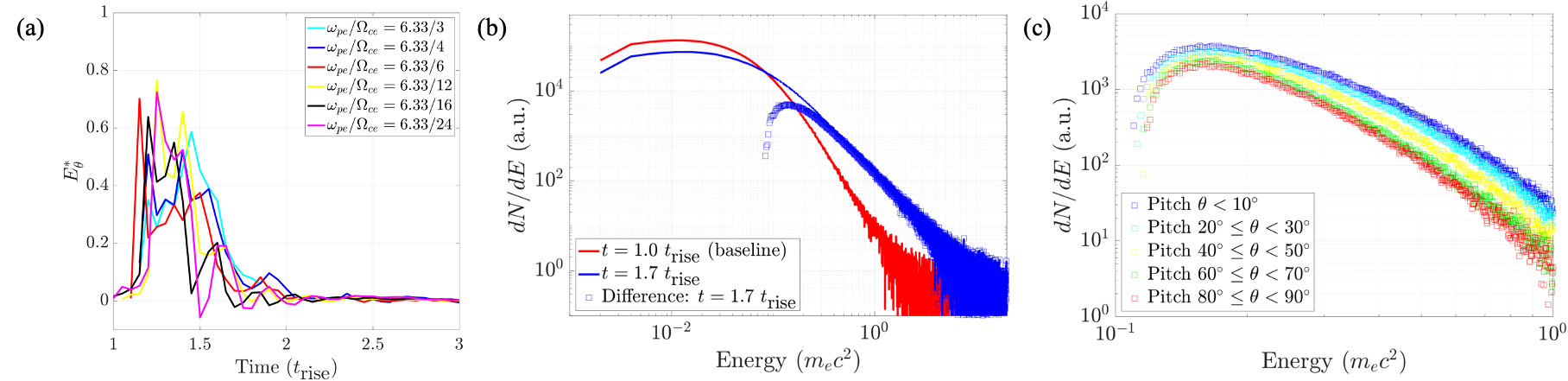}
    \caption{\small \textbf{Particle-in-cell simulations of capacitor-coil reconnection validates electron acceleration by the reconnection electric field.} \textbf{a,} The normalized electric field, or reconnection rate, $E^*_\theta$, is obtained by dividing the reconnection electric field by the upstream $V_A\times B_0$, where $V_A$ is the Alfv\'en speed computed with the upstream magnetic field $B_0$ and ion mass density in the reconnection region. From the period of $t\sim(1-2)t_{rise}$, the reconnection electric field is prominent. $E^*_\theta$ is generally constant across the $\omega_{pe}/\Omega_{ce}$ scan, representing consistent reconnection physics. \textbf{b,} The electron particle spectrum is measured within the zoomed box and with limiting the $\theta-z$ pitch angle to $10^{\circ}$. The particle spectrum for $\omega_{pe}/\Omega_{ce}=6.33/4$ is shown: $t\sim1.0t_{rise}$ (red) represents a baseline spectrum. The maximum non-thermal tail is seen at $t\sim1.7t_{rise}$, consistent with the reconnection electric field time dependence. The non-thermal difference from the baseline spectrum is shown in blue squares. \textbf{c,} Non-thermal difference spectra are shown for $t=1.7t_{rise}$, separated by $\theta-z$ pitch angle. Small pitch angles (near normal to the reconnection plane) correspond to higher-energy tails, while large pitch angles (near parallel to the reconnection plane) show smaller non-thermal acceleration.} 
  \label{fig:4}
  \end{center}
\end{figure*}

Interpretation of this particle acceleration mechanism is supported by particle-in-cell (PIC) simulations. We conducted 2-D cylindrical PIC simulations using the VPIC code~\cite{bowers08} in order to model kinetic effects and simulated particle energy spectra (Fig.~\ref{fig:3}a shows the geometric setup). The $z$-direction is the axis of symmetry, $R$ is the radial direction, and $\theta$ is the out-of-plane direction. Two rectangular-cross-section coils are placed in the simulation box, representing cross-sectional slices of the experimental U-shaped coils. Reconnection is driven by prescribing and injecting currents within the coils. We prioritize realistic mass ratio and $\beta$ in the simulation, at the expense of the reduced but scaled electron plasma frequency to electron gyrofrequency ratio, $\omega_{pe}/\Omega_{ce}$, due to the limited computational resources. Further details for the simulation setup are described in the Methods section.

The PIC simulation results demonstrate strong reconnection driven by the coil magnetic fields, with a typical out-of-plane quadrupole structure (Fig.~\ref{fig:3}c), indicative of scale separation between ions and electrons~\cite{uzdensky06,birn01}. In addition, a clear out-of-plane reconnection electric field is observed around the X-point, and the orientation of the electric field is consistent with push reconnection (see Fig.~\ref{fig:3}d).

To obtain the reconnection rate, the reconnection electric field is typically normalized by an upstream $V_{A}B_{0}$, where $B_{0}$ is the upstream magnetic field strength and $V_{A}=B_{0}/\sqrt{\mu_{0}m_{i}n_{i}}$ is the Alfv\'en velocity calculated with the ion density at the X-point. Figure~\ref{fig:4}a shows that the strongest reconnection occurs from $t\sim t_{rise}-1.7t_{rise}$, for all simulated values of $\omega_{pe}/\Omega_{ce}$. The diffusion time of the magnetic field through the plasma explains why this timing does not correspond to the expected period of push reconnection at $t < t_{rise}$. In nearly all cases, the reconnection rate reaches maximum values of $0.6-0.7$, significantly higher than the typical $\sim0.1$ rate expected for collisionless electron-ion reconnection~\cite{cassak17}: this is typical of electron-only reconnection, which is characterized by a normalized system size~\cite{pyakurel2019} $L/d_i \lesssim 5$.

Since the reconnection rate is constant across the $\omega_{pe}/\Omega_{ce}$ scan, we estimate the reconnection electric field in physical units as $E_{\textrm{rec}}\simeq0.6V_{A}B_{0}$, where the $0.6$ is the reconnection rate, as shown in Fig.~\ref{fig:4}a. Taking $B_{0}=50.7~\textrm{T}$ and a range of $n_{e}=1-5\times10^{18}~\textrm{cm}^{-3}$, $E_{\textrm{rec}}\simeq 1.3-3.0\times10^7~\textrm{V}/\textrm{m}$. This value is consistent with fitting a power law of index $k=-2.137$ to reconnection electric field strength as a function of $\omega_{pe}/\Omega_{ce}$.
\begin{table*}[htp]
\caption{Comparisons of maximum electron energy between measured values from low-$\beta$ reconnection sources, as a partial list from Ji \& Daughton (2011)~\cite{ji11}, and their estimation based on reconnection electric field acceleration, $0.1 V_A B d$. Here $d$ is a characteristic acceleration distance which is taken to the system size for maximum energy. Unless explicitly stated, plasmas consist of electrons and protons.}
\begin{center}
\resizebox{\textwidth}{!}{
\begin{tabular}{p{40mm}|c|c|c|c|c|p{60mm}}
Low-$\beta$ plasma & Size $L$ (m) & $n_e$ (m$^{-3}$) & $B$ (Tesla) & $E_\text{max,obs}$ (eV) & $E_\text{max,est}$ (eV) & Notes or assumptions \\
\hline
Laser Plasma (this work) & $1 \times 10^{-3}$ & $1\times 10^{24}$ & 50 & $(4-7)\times 10^4$ & $3\times 10^4$ & Cu$^{+18}$ plasma\\
Magnetotail\cite{oieroset02} & $6 \times 10^8$ & $1 \times 10^{5}$ & $1 \times 10^{-8}$ & $3 \times 10^{5}$ & $4 \times 10^5$ & \textit{in-situ} measurement \\
Solar Flares\cite{vilmer12,raymond12} & $1 \times 10^7$ & $1 \times 10^{15}$ & $2\times 10^{-2}$ & $1 \times 10^8$ & $6 \times 10^{10}$ & \\
X-ray Binary Disk Flares\cite{goodman08,cangemi21} & $3\times 10^4$ & $1\times 10^{24}$ & $1 \times 10^4$ & $5 \times 10^8$ & $1\times 10^{14}$ & Cygnus X-3, $M=10M_\odot$, $R=R_S$\\
Crab Nebula Flares\cite{kroon16,tavani11,abdo11} & $1 \times 10^{17}$ & $10^6$ & $1 \times 10^{-8}$ & $5\times 10^{15}$ & $2.4\times 10^{15}$& pair plasma \\
Gamma Ray Bursts\cite{sari99,uzdensky11} & $10^4$ & $ 2\times 10^{35}$ & $4 \times 10^9$ & $5\times 10^{9}$ & $3 \times 10^{20}$ & pair plasma\\
Magnetar Flares\cite{beloborodov17} &  $10^4$ & $10^{41}$ & $2 \times 10^{11}$ & $2 \times 10^{8}$ & $5 \times 10^{20}$ & pair plasma, FRB 121102\\
AGN Disk Flares\cite{torricelli05,goodman08} & $3\times 10^{11}$ & $1\times 10^{17}$ & 4 & $5 \times10^{8}$ & $3\times 10^{17}$ &Seyfert 1 NGC 5548, $M=10^8M_\odot$, $R=R_S$\\
Radio Lobes\cite{massaro11} & $3\times 10^{19}$ & 0.1 & $5\times 10^{-10}$ & $5 \times 10^{11} $ & $5 \times 10^{16} $ & \\
Extragalactic Jets\cite{kataoka03} & $3 \times 10^{19}$ & $3 \times 10^1$ & $10^{-7}$ & $7 \times 10^{12}$ & $1 \times 10^{18}$ & 3C 303\\
\end{tabular}}
\end{center}
\label{astro-table}
\end{table*}%

\begin{figure*}[htb!]
  \begin{center}
    \includegraphics[width=\textwidth]{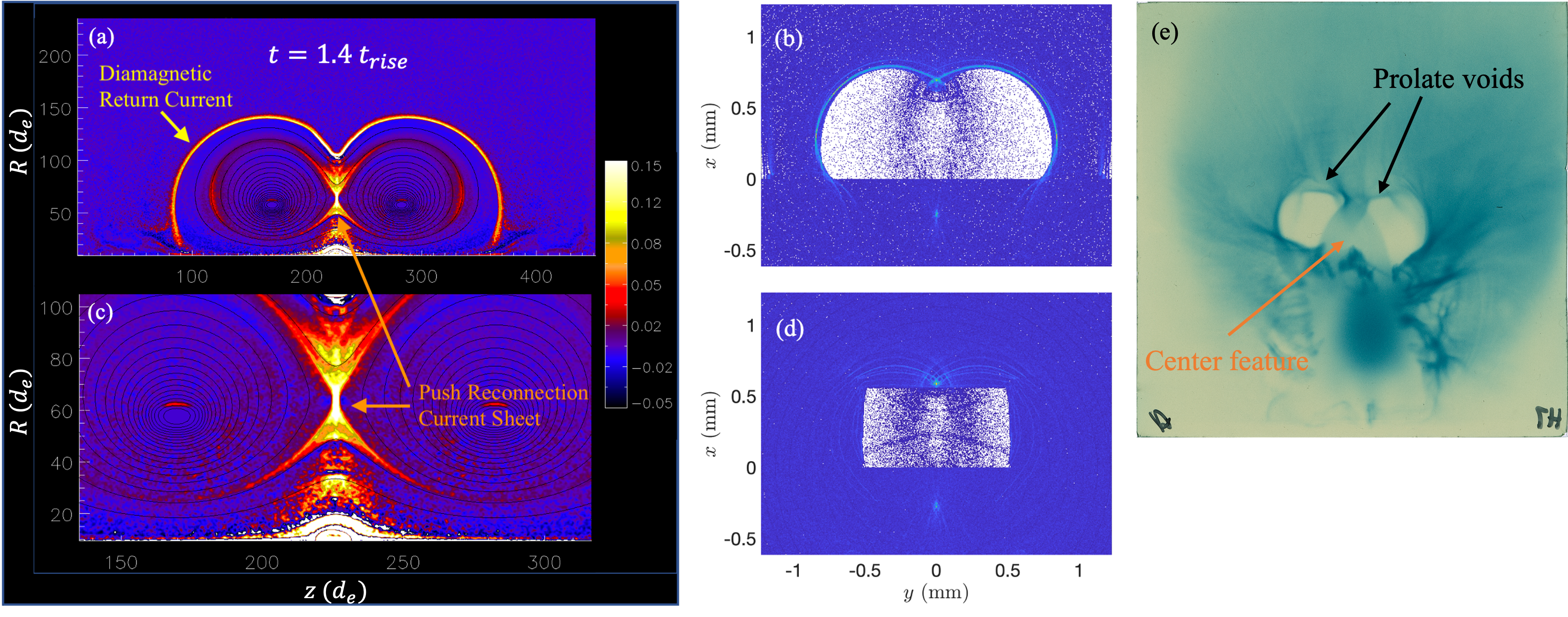}
  \end{center}
  \caption{\textbf{The ``center feature'' in experimental proton radiographs is reproduced by synthetic raytracing of PIC-generated fields, and likely formed from the push reconnection current sheet.} \textbf{a,} Full box and \textbf{c,} zoomed out-of-plane current $j_{\theta}$ profiles with respective synthetic proton radiographs (\textbf{b, d}) for $t=1.4~t_{\textrm{rise}}$. The synthetic proton radiographs are generated by ``sweeping'' the electromagnetic field structure in a semicircle, advancing a cone of protons through the fields in the $z$ direction, and recording the projected proton positions in the $y$-$x$ plane. The diamagnetic return current can be seen in the full $j_{\theta}$ profile, but not in the zoomed profile. \textbf{e,} Experimental proton radiograph taken at $t=1.0~\textrm{ns}$, proton energy $E_{p}=18.3~\textrm{MeV}$. Two primary features are seen: two prolate voids are caused by the coil magnetic fields, and the center feature can be explained by electromagnetic fields corresponding to a push reconnection current sheet. In the synthetic radiographs, white areas represent no proton fluence, dark blue areas represent the background proton level, and bright blue areas represent the highest proton concentration. In the experimental radiograph, white areas represent no proton fluence, and darker shades of green represent larger proton concentrations.}
  \label{fig:vpic_zoom_1.4}
\end{figure*}

Electromagnetic fields from the PIC simulations are analyzed through a synthetic proton raytracing algorithm to predict a dark center feature corresponding to the reconnection current sheet during push reconnection. This center feature is observed in experimental proton radiographs taken at $t=1.0~\textrm{ns}$ after the laser pulse (Fig.~\ref{fig:vpic_zoom_1.4}e), indicating the presence of reconnection in the experimental platform. To generate the synthetic proton radiograph, protons with kinetic energy of $50~\textrm{MeV}$ are advanced via a 4th-order Runge-Kutta algorithm. At each proton position, electromagnetic fields are inferred from the 2-D PIC fields, ``swept'' in angle along the semi-circular portion of the coils. More details of the synthetic raytracing algorithm are described in the Methods section.

Synthetic raytracing is performed for $t=1.4~t_{\textrm{rise}}$, corresponding to strong push reconnection. The center feature is reproduced in the radiograph (Fig.~\ref{fig:vpic_zoom_1.4}b), and two primary features in the out-of-plane current $j_{\theta}$ profiles (Fig.~\ref{fig:vpic_zoom_1.4}a) are potentially responsible for creating this center feature: the push reconnection current sheet and diamagnetic return current. To de-convolve the effects of each on the synthetic proton radiographs, raytracing is performed on a ``zoomed'' field, where the diamagnetic return current is largely shielded out (Fig.~\ref{fig:vpic_zoom_1.4}c). The center feature is maintained in this radiograph (Fig.~\ref{fig:vpic_zoom_1.4}d), indicating the source of the center feature as the push reconnection current sheet. Thus, the presence of a similar center feature in experimental radiographs is indicative of push reconnection and the corresponding electromagnetic fields.

Finally, the PIC simulations demonstrate non-thermal particle acceleration during the push phase of reconnection. Various filters are applied to the electron population to select the electrons that best compare to the experimental  spectra. First, to focus on the electrons that are affected by reconnection, electrons are  measured only within the zoomed-in simulation area shown in Fig.~\ref{fig:3}a. Second,  the $\theta-z$ pitch angle is  limited  to select electrons that can escape and be measured by the OU-ESM detector. For our 2-D axisymmetric simulation, we do not limit the $\theta-R$ pitch angle, since 
for any fixed detector angle and $\theta-R$ pitch angle, a reconnection plane exists such that a particle accelerated from that plane would reach the detector.

Due to the particle injection scheme from $0<t<t_{\textrm{rise}}$, the baseline spectrum is taken at $t=t_{\textrm{rise}}$, in order to distinguish reconnection-accelerated electrons from injected electrons. The reconnection rate evolution shows that reconnection does not begin until $t>t_{\textrm{rise}}$, further validating this approach. Figure~\ref{fig:4}b shows the formation of a non-thermal electron tail. The tail grows larger with time, up to a maximum (at $t\sim1.7t_{rise}$), and begins to decay back to a Maxwellian, as reconnection stops and accelerated particles escape the system through the open boundaries. The time of the maximally non-thermal spectrum corresponds well to the reconnection electric field time dependence, demonstrating push reconnection as the source of the accelerated particles.

Comparison of the experimental particle spectra with PIC simulations supports acceleration by the reconnection electric field as the primary acceleration mechanism that forms the non-thermal electron tail. This is evidenced by the angular dependence and accelerated energies of the non-thermal tails in experimental measurements. The strongest non-thermal components are seen in Channel 5, corresponding to its near-normal orientation. The strength of the bump decreases as the azimuthal angle grows more oblique. Acceleration by the out-of-plane reconnection electric field would be expected to produce this angular dependence: electrons with larger pitch angles would be directed into regions with high field, resulting in re-magnetization, preventing the electron from reaching the detector. This angular dependence of the accelerated electrons is confirmed by PIC simulations, as shown in Fig.~\ref{fig:4}c, where non-thermal electrons decrease with increasing pitch angle from the reconnection electric field direction. 

Other proposed acceleration mechanisms are not expected to be applicable because the required conditions for them are not satisfied in our experiment. Fermi acceleration typically requires multiple plasmoids in the current sheet as acceleration sites. Parallel electric field acceleration requires a finite guide field. Betatron acceleration requires increasing magnetic field in the downstream region. Polarization drift acceleration is not significant for electrons.

Although the accelerated particle spectrum from simulations could not be obtained for the experimental value of $\omega_{pe}/\Omega_{ce}=6.33$ through scaling due to prohibitive computational cost (see Methods section), the simulation-determined scaling of the out-of-plane electric field is well-established. Using the calculated reconnection electric field $E_{\theta}=(1.3-3.0)\times{10}^7~\textrm{V}/\textrm{m}$, a simple estimate for the expected accelerated electron energy gain becomes $\Delta E\sim \left| q_{e}\right| E_{\theta} d$, 
where $q_{e}$ is the electron charge, and $d$ is a characteristic acceleration distance, here taken to be $d\sim1000~\mu\textrm{m}$. This predicts $13-30~\textrm{keV}$ electrons, which represents an upper bound on accelerated electron energy with this mechanism, and is within a factor of 2 of the experimental bump of $\sim50-70~\textrm{keV}$. Several potential factors can explain this discrepancy. First, a larger reconnection electric field and thus larger electron acceleration can be achieved with a larger than expected upstream magnetic field or a smaller than expected plasma density since $E_\theta \propto V_A$. The former can occur due to magnetic field pileup in the upstream region. Plasma density near the X-point is uncertain because the Thomson scattering probes the plasma located in the downstream region above the coils.

Second, there is a possibility that the bump may not be due to reconnection: instead, due to the different magnetic topologies of the reconnection and one-coil cases, LPI-generated electrons of certain energies may be preferentially deflected toward certain angles, contributing to the observed spectral features. In the Supplementary Material section, this possibility is analyzed in detail using electron/positron raytracing with vacuum magnetic fields, and we conclude that this coil magnetic field deflection alone is unable to produce the experimental spectral trends and features observed. Plasma effects in the raytracing simulations are expected to be small, as illustrated by the low signal level in the no-coil electron spectra (Fig.~\ref{fig:2}f). By eliminating this alternative interpretation, this further supports that the detected electron spectral bump is due to reconnection.

The inference that direct electric field acceleration is operating in the experiments motivates estimating the corresponding attainable particle energies from this mechanism in representative low-$\beta$ collisionless reconnecting plasmas throughout the Universe~\cite{ji11} and comparing to maximum inferred electron energies from observations. The result is  shown in Table 1, where we have assumed that our experimental implications for the mostly electron-only reconnection regime can be extended to electron-ion or pair plasma reconnection regimes. This  leads to the reconnection electric field $E_\theta = 0.1 V_A B$ typically found in collisionless reconnection~\cite{cassak17}. Therefore, the upper bound for the energy of the accelerated electrons by the reconnection electric field is established by the Hillas limit~\cite{hillas_1984} $E_{\text{max,est}}=eE_\theta d$, where $d$ is a characteristic acceleration distance, here taken to be the system size $L$. 

The estimated maximum energy is within a factor of 2 for Earth's magnetotail and Crab nebula flares, implying that if this mechanism is responsible for acceleration of the most energetic electrons in these two cases, coherent acceleration over a distance comparable to the system size is required. In all other cases, the observed maximum electron energy is well below the estimated theoretical maximum energy, suggesting that if this mechanism is at work, it must operate over length scales much shorter than the system size but with a properly distributed spread to populate the whole electron energy spectrum. Interestingly, this scaling of maximum energy has been also identified in large-scale simulations at low-$\beta$ but in relativistic regimes~\cite{guo15}.

Our laser-powered capacitor coils offer a unique experimental reconnection platform in magnetically driven low-$\beta$ plasmas to further study acceleration of electrons (and ions) in various reconnection regimes~\cite{ji11,ji22} via direct detection of accelerated particles. The extent to which the same or different mechanisms~\cite{blandford17,guo20} of particle acceleration emerge in different regimes will be of great interest to determine in future laboratory research and may depend on the particular reconnection boundary conditions and system geometries. Although no other mechanisms are excluded, our reported results serve as the first direct evidence of any hypothesized acceleration mechanism, by magnetic reconnection.

\section*{Methods}
\subsection*{Laser-powered capacitor coil target}
Laser-powered capacitor coil targets are composed of parallel plates (the capacitor) connected by one or multiple wires (the coils). Holes are formed in the front (facing the driving laser) plate to allow the laser beam(s) to bypass the front plate and only hit the back plate. Superthermal hot electrons are generated during the intense laser-solid interaction, some of which manage to escape from the back plate. A strong current is therefore supplied to the U-shaped wires from back to front, due to the resultant potential difference between the plates.

We use capacitor coil targets made from $50~\mu\textrm{m}$-thick copper. The capacitors are  formed by two square parallel plates with length $1.5~\textrm{mm}$, with an inter-plate distance of $600~\mu\textrm{m}$. Two holes of radius $250~\mu\textrm{m}$ are formed in the front plate to accommodate OMEGA EP long-pulse beams 3 and 4. The plates are joined by one or two parallel U-shaped coils, with rectangular cross-section $50~\mu\textrm{m}\times100~\mu\textrm{m}$. Each coil consists of two $500~\mu\textrm{m}$ straight sections, joined by a semicircular section with radius $300~\mu\textrm{m}$. In two-coil targets, the coils are separated by $600~\mu\textrm{m}$. Capacitor coil targets are fabricated by laser-cutting a design in $50~\mu\textrm{m}$-thick sheet copper, then bending the coils into shape.

\subsection*{Particle spectra measurement using OU-ESM}
The Osaka University electron spectrometer~\cite{habara_2019_ouesm} is a time-integrated diagnostic that can provide angular resolution in either polar or azimuthal angles, relative to the target. This is accomplished by the use of 5 channels, each separated in angle by $5^{\circ}$. In the experiment, we chose the azimuthal angle spread, since this pitch angle allows for distinguishing between acceleration mechanisms in an axisymmetric setup, with symmetry in the polar direction. 

After reaching the spectrometer, an electron first passes through a pinhole $700~\mu\textrm{m}$ wide and $2~\textrm{cm}$ deep. Separation of electron energies is accomplished with a set of permanent magnets placed along the detector line-of-sight, creating a magnetic field perpendicular to the line-of-sight. The $\vec{v}\times\vec{B}$ force deflects differently-energized electrons different distances along the detector length onto a BAS-TR image plate. In general, impacts closer to the detector entrance represent lower-energy electrons. In the experiment, magnets were chosen corresponding to electron energies in the $20~\textrm{keV} - 1~\textrm{MeV}$ range.

The field of view and solid angle subtended by each channel are defined by the pinhole size $p=700~\mu\textrm{m}$, distance to target $D=37.5~\textrm{cm}$, and pinhole/collimator depth $d=2~\textrm{cm}$. The field of view is given by $\textrm{FOV}=pD/d=13.1~\textrm{mm}$, which is significantly larger than the target size and sufficient to capture electrons from the main interaction. The solid angle is $\Omega=P^{2}/D^{2}=3.5\times10^{-6}~\textrm{sr}$ and provides the primary restriction for electrons reaching the detector.

The primary sources of error in interpreting OU-ESM data involve image plate response to energetic electrons and image plate scanning offsets. The image plate response is taken from Bonnet et al., 2013~\cite{bonnet_2013_ip}, and introduces a $28\%$ uncertainty. In addition, image plate signals decay with time, so image plates are scanned at exactly $30$ minutes after the shot for consistency in signal level. In interpreting the image plate, defining the edge of the magnets is critical to an accurate energy spectrum. Here, an uncertainty of approximately $5$ pixels, or $0.5~\textrm{mm}$ is introduced, translating to an uncertainty in the spectrum energy.

\subsection*{Particle-in-cell simulation setup}
The 2-D PIC simulation box has dimensions of $L_{z} = 4D_{\textrm{coil}}$ and $L_{R}=2D_{\textrm{coil}}$ in the $z$ and $R$ directions, respectively, where $D_{\textrm{coil}}$ is the inter-coil distance of $L=600~\mu\textrm{m}$. To avoid the difficult boundary at $R=0$, a minimum radius of $R_{\textrm{min}}=50~\mu\textrm{m}$ was used. Two rectangular-cross-section coils of width $l_{z}=100~\mu\textrm{m}$ and height $l_{R}=50~\mu\textrm{m}$ are located at $R=300~\mu\textrm{m}$ and $z=\pm D_{\textrm{coil}}/2=\pm300~\mu\textrm{m}$, matching experimental positions.

In the simulation, lengths are normalized to electron skin depth $d_{e}=c/\omega_{pe}$, and times are expressed in terms of the inverse electron plasma frequency $\omega_{pe}^{-1}$. Due to the large Lundquist number, collisions are turned off in the simulation.

It is computationally untenable to perform a simulation with completely  physical parameters, and so  priorities must be made. An accurate particle spectrum is of great importance, so reducing ion-to-electron mass ratio $m_{i}/m_{e}$ is undesirable. In addition, the plasma $\beta$ has been shown to be a critical parameter in particle acceleration~\cite{Li_2018}, so we maintain the physical $\beta$ in the simulation. To reduce computational time, we instead use artificially small values of the electron plasma frequency to electron gyrofrequency  $\omega_{pe}/\Omega_{ce}$.   By keeping $\beta$ constant, a reduced $\omega_{pe}/\Omega_{ce}$ represents an artificially strong magnetic field, coupled to an artificially hot plasma. Scaling relations for electromagnetic field strength can be established as a function of $\omega_{pe}/\Omega_{ce}$ in order to extrapolate to physical conditions. The physical $\omega_{pe}/\Omega_{ce}=6.33$, and simulations are run for reduced values $\omega_{pe}/\Omega_{ce}=6.33/3$, $6.33/4$, $6.33/6$, $6.33/12$, $6.33/16$, and $6.33/24$.

Cell size is limited by the Debye length, so the number of cells changes with $\omega_{pe}/\Omega_{ce}$. At $\omega_{pe}/\Omega_{ce}=6.33/4$, the number of cells is $n_{z}\times n_{R}=1440\times720$, spanning $L_{z}\times L_{R}=451.6d_{e}\times225.8d_{e}$. $200$ macro-particles of each species are initialized per cell. The achievable cell size and number of macro-particles per cell also limit a viable scaling of accelerated electron spectra to be established for the small electron diffusion region where electrons are demagnetized, and thus are free to be accelerated by the reconnection electric field.

For physical $\omega_{pe}/\Omega_{ce}=6.33$, the simulation is initialized with a uniform Maxwellian plasma with $n_{e}=10^{18}~\textrm{cm}^{-3}, n_{i}=n_{e}/Z=5.6\times10^{16}~\textrm{cm}^{-3}$, $T_{e}=T_{i}=400~\textrm{eV}$ to match experimental parameters. Compared to experiment, a lower initial plasma density is used due to the inclusion of a particle injection scheme from the coil region. A representative magnetic field strength $B_{0}=50.7~\textrm{T}$ is taken to be the upstream magnetic field at $z=-D_{\textrm{coil}}/4$ from the center between the coils. The simulation $\beta$ is therefore $0.063$. Due to the artificially reduced $\omega_{pe}/\Omega_{ce}$ values, ion and electron temperatures and coil magnetic fields are artificially increased, while keeping density constant to maintain plasma $\beta$.

Electrically conducting boundary conditions are set for fields, and open boundary conditions~\cite{daughton_openbc_2006} are set for particles in the $z$ and $R$ directions (periodic boundary conditions are set for $\theta$). The open boundary conditions prevent accelerated particles that would otherwise escape the system from being re-accelerated. Our choice therefore prevents an over-estimation of particle acceleration that would be inevitable with periodic or reflecting boundaries.

At $t=0$, electromagnetic fields are set to 0. The capacitor coil currents are modeled by injecting currents with the following time profile:
\begin{equation}
    I_{\textrm{coil}}(t)=
    \begin{cases}
    I_{0}(t/t_{\textrm{rise}}), &t<t_{\textrm{rise}}\\
    I_{0}\exp(-(t-t_{\textrm{rise}})/t_{\textrm{decay}}), &t\ge t_{\textrm{rise}}
    \end{cases}
\end{equation}
where $t_{\textrm{rise}}=1~\textrm{ns}$, $t_{\textrm{decay}}=8.6~\textrm{ns}$, and $I_{0}=57~\textrm{kA}$  match experimental measurements. Currents are oriented into the page ($-\theta$ direction). The coil magnetic fields are then calculated from the current distribution within the coils.

The reconnection plasma primarily emanates from the coils, due to ablation of copper plasma by Ohmic heating within the coils and irradiation by x-rays from the laser interaction. In contrast, the plasma generated at the laser spot takes a few nanoseconds to flow into the reconnection region, and  does not play a significant role in reconnection, particularly during the push phase. To simulate the coil plasma, we use a particle injection scheme: a volume injector is implemented around the coils, with a Gaussian spatial profile (Gaussian width and height are set to $l_{z}$ and $l_{R}$, respectively), and a linear time-dependence from $t=0-1~\textrm{ns}$. The injection rate is tuned to match experimental density measurements. Without the particle injection scheme, density voids form around the coils, as the strong magnetic field pressure pushes out plasma as the magnetic field diffuses outwards from the coils.

\subsection*{Proton raytracing using PIC electromagnetic fields}

Protons are advanced through a 3-D representation of the PIC fields through a 4th-order Runge-Kutta algorithm. Electric and magnetic fields are applied onto a 3-D grid by projecting from a 2-D simulation using the following methodology:
\begin{enumerate}
    \item The center of the semi-circular coil is defined at the origin. For each proton near the coil region, the radius and azimuthal angle relative to the top-most point of the coil are calculated ($\theta=0$ corresponds to the ``vertical'' reconnection plane).
    \item For angles corresponding to the lower hemisphere ($\theta>90^{\circ}$ or $\theta<-90^{\circ}$), electromagnetic fields are assumed to be $0$.
    \item If the proton is far away enough from the coils to be outside the effective PIC simulation box ($r>R_{\textrm{PIC,max}}$ or $r<R_{\textrm{PIC,min}}$), electromagnetic fields are assumed to be $0$.
    \item The 3-D electric and magnetic fields corresponding to the radial and axial positions in the reconnection plane are found with linear interpolation.
    \item These fields are rotated by the azimuthal angle that corresponds to the proton location.
\end{enumerate}

Simply, the 2-D reconnection plane is ``swept'' in angle along the semi-circular portion of the coils. For each synthetic radiograph, $10^{7}$ protons are sampled, with a maximum source angle of $0.4~\textrm{rad}$. The ``impact coordinates'' of each proton at a pre-defined synthetic detector are combined into a 2-D histogram, spanning $L_{y}=L_{x}=2.46~\textrm{mm}$ in both directions with $500$ bins in each direction. Each bin thus represents a $4.9~\mu\textrm{m}$ width.

\acknowledgments
This work was supported by the LaserNetUS program and the High Energy Density Laboratory Plasma Science program by Office of Science, Fusion Energy Sciences (FES) and NNSA under Grant No. DE-SC0020103. The authors express their gratitude to General Atomics, the University of Michigan, and the Laboratory for Laser Energetics (LLE) for target fabrication, and to the OMEGA and OMEGA EP crews for experimental and technical support. The work was also supported by DOE Grant GR523126 and NSF Grant PHY-2020249.

\bibliography{reconnection,1MAIN,X-ray}

\begin{thebibliography}{65}%
\makeatletter
\providecommand \@ifxundefined [1]{%
 \@ifx{#1\undefined}
}%
\providecommand \@ifnum [1]{%
 \ifnum #1\expandafter \@firstoftwo
 \else \expandafter \@secondoftwo
 \fi
}%
\providecommand \@ifx [1]{%
 \ifx #1\expandafter \@firstoftwo
 \else \expandafter \@secondoftwo
 \fi
}%
\providecommand \natexlab [1]{#1}%
\providecommand \enquote  [1]{``#1''}%
\providecommand \bibnamefont  [1]{#1}%
\providecommand \bibfnamefont [1]{#1}%
\providecommand \citenamefont [1]{#1}%
\providecommand \href@noop [0]{\@secondoftwo}%
\providecommand \href [0]{\begingroup \@sanitize@url \@href}%
\providecommand \@href[1]{\@@startlink{#1}\@@href}%
\providecommand \@@href[1]{\endgroup#1\@@endlink}%
\providecommand \@sanitize@url [0]{\catcode `\\12\catcode `\$12\catcode
  `\&12\catcode `\#12\catcode `\^12\catcode `\_12\catcode `\%12\relax}%
\providecommand \@@startlink[1]{}%
\providecommand \@@endlink[0]{}%
\providecommand \url  [0]{\begingroup\@sanitize@url \@url }%
\providecommand \@url [1]{\endgroup\@href {#1}{\urlprefix }}%
\providecommand \urlprefix  [0]{URL }%
\providecommand \Eprint [0]{\href }%
\providecommand \doibase [0]{http://dx.doi.org/}%
\providecommand \selectlanguage [0]{\@gobble}%
\providecommand \bibinfo  [0]{\@secondoftwo}%
\providecommand \bibfield  [0]{\@secondoftwo}%
\providecommand \translation [1]{[#1]}%
\providecommand \BibitemOpen [0]{}%
\providecommand \bibitemStop [0]{}%
\providecommand \bibitemNoStop [0]{.\EOS\space}%
\providecommand \EOS [0]{\spacefactor3000\relax}%
\providecommand \BibitemShut  [1]{\csname bibitem#1\endcsname}%
\let\auto@bib@innerbib\@empty
\bibitem [{\citenamefont {Yamada}, \citenamefont {Kulsrud},\ and\ \citenamefont
  {Ji}(2010)}]{yamada10}%
  \BibitemOpen
  \bibfield  {author} {\bibinfo {author} {\bibfnamefont {M.}~\bibnamefont
  {Yamada}}, \bibinfo {author} {\bibfnamefont {R.}~\bibnamefont {Kulsrud}}, \
  and\ \bibinfo {author} {\bibfnamefont {H.}~\bibnamefont {Ji}},\ }\href@noop
  {} {\bibfield  {journal} {\bibinfo  {journal} {Rev. Mod. Phys.}\ }\textbf
  {\bibinfo {volume} {82}},\ \bibinfo {pages} {603} (\bibinfo {year}
  {2010})}\BibitemShut {NoStop}%
\bibitem [{\citenamefont {Ji}\ \emph {et~al.}(2022)\citenamefont {Ji},
  \citenamefont {Daughton}, \citenamefont {Jara-Almonte}, \citenamefont {Le},
  \citenamefont {Stanier},\ and\ \citenamefont {Yoo}}]{ji22}%
  \BibitemOpen
  \bibfield  {author} {\bibinfo {author} {\bibfnamefont {H.}~\bibnamefont
  {Ji}}, \bibinfo {author} {\bibfnamefont {W.}~\bibnamefont {Daughton}},
  \bibinfo {author} {\bibfnamefont {J.}~\bibnamefont {Jara-Almonte}}, \bibinfo
  {author} {\bibfnamefont {A.}~\bibnamefont {Le}}, \bibinfo {author}
  {\bibfnamefont {A.}~\bibnamefont {Stanier}}, \ and\ \bibinfo {author}
  {\bibfnamefont {J.}~\bibnamefont {Yoo}},\ }\href@noop {} {\enquote {\bibinfo
  {title} {Magnetic reconnection in the era of exascale computing and
  multiscale experiments},}\ }\bibinfo {howpublished} {in press, Nat. Rev.
  Phys.} (\bibinfo {year} {2022})\BibitemShut {NoStop}%
\bibitem [{\citenamefont {{Blandford}}\ \emph {et~al.}(2017)\citenamefont
  {{Blandford}}, \citenamefont {{Yuan}}, \citenamefont {{Hoshino}},\ and\
  \citenamefont {{Sironi}}}]{blandford17}%
  \BibitemOpen
  \bibfield  {author} {\bibinfo {author} {\bibfnamefont {R.}~\bibnamefont
  {{Blandford}}}, \bibinfo {author} {\bibfnamefont {Y.}~\bibnamefont {{Yuan}}},
  \bibinfo {author} {\bibfnamefont {M.}~\bibnamefont {{Hoshino}}}, \ and\
  \bibinfo {author} {\bibfnamefont {L.}~\bibnamefont {{Sironi}}},\ }\href@noop
  {} {\bibfield  {journal} {\bibinfo  {journal} {Space Science Rev.}\ }\textbf
  {\bibinfo {volume} {207}},\ \bibinfo {pages} {291} (\bibinfo {year}
  {2017})}\BibitemShut {NoStop}%
\bibitem [{\citenamefont {Guo}\ \emph {et~al.}(2020)\citenamefont {Guo},
  \citenamefont {Liu}, \citenamefont {Li}, \citenamefont {Li}, \citenamefont
  {Daughton},\ and\ \citenamefont {Kilian}}]{guo20}%
  \BibitemOpen
  \bibfield  {author} {\bibinfo {author} {\bibfnamefont {F.}~\bibnamefont
  {Guo}}, \bibinfo {author} {\bibfnamefont {Y.-H.}\ \bibnamefont {Liu}},
  \bibinfo {author} {\bibfnamefont {X.}~\bibnamefont {Li}}, \bibinfo {author}
  {\bibfnamefont {H.}~\bibnamefont {Li}}, \bibinfo {author} {\bibfnamefont
  {W.}~\bibnamefont {Daughton}}, \ and\ \bibinfo {author} {\bibfnamefont
  {P.}~\bibnamefont {Kilian}},\ }\href@noop {} {\bibfield  {journal} {\bibinfo
  {journal} {Phys. Plasmas}\ }\textbf {\bibinfo {volume} {27}},\ \bibinfo
  {pages} {080501} (\bibinfo {year} {2020})}\BibitemShut {NoStop}%
\bibitem [{\citenamefont {Masuda}\ \emph {et~al.}(1994)\citenamefont {Masuda},
  \citenamefont {Kosugi}, \citenamefont {Hara}, \citenamefont {Tsuneta},\ and\
  \citenamefont {Ogawara}}]{masuda94}%
  \BibitemOpen
  \bibfield  {author} {\bibinfo {author} {\bibfnamefont {S.}~\bibnamefont
  {Masuda}}, \bibinfo {author} {\bibfnamefont {T.}~\bibnamefont {Kosugi}},
  \bibinfo {author} {\bibfnamefont {H.}~\bibnamefont {Hara}}, \bibinfo {author}
  {\bibfnamefont {S.}~\bibnamefont {Tsuneta}}, \ and\ \bibinfo {author}
  {\bibfnamefont {Y.}~\bibnamefont {Ogawara}},\ }\href@noop {} {\bibfield
  {journal} {\bibinfo  {journal} {Nature}\ }\textbf {\bibinfo {volume} {371}},\
  \bibinfo {pages} {495} (\bibinfo {year} {1994})}\BibitemShut {NoStop}%
\bibitem [{\citenamefont {{{\O}ieroset}}\ \emph {et~al.}(2002)\citenamefont
  {{{\O}ieroset}}, \citenamefont {{Lin}}, \citenamefont {{Phan}}, \citenamefont
  {{Larson}},\ and\ \citenamefont {{Bale}}}]{oieroset02}%
  \BibitemOpen
  \bibfield  {author} {\bibinfo {author} {\bibfnamefont {M.}~\bibnamefont
  {{{\O}ieroset}}}, \bibinfo {author} {\bibfnamefont {R.~P.}\ \bibnamefont
  {{Lin}}}, \bibinfo {author} {\bibfnamefont {T.~D.}\ \bibnamefont {{Phan}}},
  \bibinfo {author} {\bibfnamefont {D.~E.}\ \bibnamefont {{Larson}}}, \ and\
  \bibinfo {author} {\bibfnamefont {S.~D.}\ \bibnamefont {{Bale}}},\
  }\href@noop {} {\bibfield  {journal} {\bibinfo  {journal} {Phys. Rev. Lett.}\
  }\textbf {\bibinfo {volume} {89}},\ \bibinfo {pages} {195001} (\bibinfo
  {year} {2002})}\BibitemShut {NoStop}%
\bibitem [{\citenamefont {{Krucker}}\ \emph {et~al.}(2010)\citenamefont
  {{Krucker}}, \citenamefont {{Hudson}}, \citenamefont {{Glesener}},
  \citenamefont {{White}}, \citenamefont {{Masuda}}, \citenamefont
  {{Wuelser}},\ and\ \citenamefont {{Lin}}}]{krucker10}%
  \BibitemOpen
  \bibfield  {author} {\bibinfo {author} {\bibfnamefont {S.}~\bibnamefont
  {{Krucker}}}, \bibinfo {author} {\bibfnamefont {H.~S.}\ \bibnamefont
  {{Hudson}}}, \bibinfo {author} {\bibfnamefont {L.}~\bibnamefont
  {{Glesener}}}, \bibinfo {author} {\bibfnamefont {S.~M.}\ \bibnamefont
  {{White}}}, \bibinfo {author} {\bibfnamefont {S.}~\bibnamefont {{Masuda}}},
  \bibinfo {author} {\bibfnamefont {J.-P.}\ \bibnamefont {{Wuelser}}}, \ and\
  \bibinfo {author} {\bibfnamefont {R.~P.}\ \bibnamefont {{Lin}}},\ }\href@noop
  {} {\bibfield  {journal} {\bibinfo  {journal} {Astrophys. J.}\ }\textbf
  {\bibinfo {volume} {714}},\ \bibinfo {pages} {1108} (\bibinfo {year}
  {2010})}\BibitemShut {NoStop}%
\bibitem [{\citenamefont {Abdo}\ \emph {et~al.}(2011)\citenamefont {Abdo},
  \citenamefont {Ackermann}, \citenamefont {Ajello}, \citenamefont {Allafort},
  \citenamefont {Baldini}, \citenamefont {Ballet}, \citenamefont {Barbiellini},
  \citenamefont {Bastieri}, \citenamefont {Bechtol},\ and\ \citenamefont
  {Bellazzini}}]{abdo11}%
  \BibitemOpen
  \bibfield  {author} {\bibinfo {author} {\bibfnamefont {A.}~\bibnamefont
  {Abdo}}, \bibinfo {author} {\bibfnamefont {M.}~\bibnamefont {Ackermann}},
  \bibinfo {author} {\bibfnamefont {M.}~\bibnamefont {Ajello}}, \bibinfo
  {author} {\bibfnamefont {A.}~\bibnamefont {Allafort}}, \bibinfo {author}
  {\bibfnamefont {L.}~\bibnamefont {Baldini}}, \bibinfo {author} {\bibfnamefont
  {J.}~\bibnamefont {Ballet}}, \bibinfo {author} {\bibfnamefont
  {G.}~\bibnamefont {Barbiellini}}, \bibinfo {author} {\bibfnamefont
  {D.}~\bibnamefont {Bastieri}}, \bibinfo {author} {\bibfnamefont
  {K.}~\bibnamefont {Bechtol}}, \ and\ \bibinfo {author} {\bibfnamefont
  {R.}~\bibnamefont {Bellazzini}},\ }\href@noop {} {\bibfield  {journal}
  {\bibinfo  {journal} {Science}\ }\textbf {\bibinfo {volume} {331}},\ \bibinfo
  {pages} {739} (\bibinfo {year} {2011})}\BibitemShut {NoStop}%
\bibitem [{\citenamefont {Tavani}\ \emph {et~al.}(2011)\citenamefont {Tavani},
  \citenamefont {Bulgarelli}, \citenamefont {Vittorini}, \citenamefont
  {Pellizzoni}, \citenamefont {Striani}, \citenamefont {Caraveo}, \citenamefont
  {Weisskopf}, \citenamefont {Tennant}, \citenamefont {Pucella},\ and\
  \citenamefont {Trois}}]{tavani11}%
  \BibitemOpen
  \bibfield  {author} {\bibinfo {author} {\bibfnamefont {M.}~\bibnamefont
  {Tavani}}, \bibinfo {author} {\bibfnamefont {A.}~\bibnamefont {Bulgarelli}},
  \bibinfo {author} {\bibfnamefont {V.}~\bibnamefont {Vittorini}}, \bibinfo
  {author} {\bibfnamefont {A.}~\bibnamefont {Pellizzoni}}, \bibinfo {author}
  {\bibfnamefont {E.}~\bibnamefont {Striani}}, \bibinfo {author} {\bibfnamefont
  {P.}~\bibnamefont {Caraveo}}, \bibinfo {author} {\bibfnamefont
  {M.}~\bibnamefont {Weisskopf}}, \bibinfo {author} {\bibfnamefont
  {A.}~\bibnamefont {Tennant}}, \bibinfo {author} {\bibfnamefont
  {G.}~\bibnamefont {Pucella}}, \ and\ \bibinfo {author} {\bibfnamefont
  {A.}~\bibnamefont {Trois}},\ }\href@noop {} {\bibfield  {journal} {\bibinfo
  {journal} {Science}\ }\textbf {\bibinfo {volume} {331}},\ \bibinfo {pages}
  {736} (\bibinfo {year} {2011})}\BibitemShut {NoStop}%
\bibitem [{\citenamefont {{Chen}}\ \emph {et~al.}(2020)\citenamefont {{Chen}},
  \citenamefont {{Shen}}, \citenamefont {{Gary}}, \citenamefont {{Reeves}},
  \citenamefont {{Fleishman}}, \citenamefont {{Yu}}, \citenamefont {{Guo}},
  \citenamefont {{Krucker}}, \citenamefont {{Lin}}, \citenamefont {{Nita}},\
  and\ \citenamefont {{Kong}}}]{chen20}%
  \BibitemOpen
  \bibfield  {author} {\bibinfo {author} {\bibfnamefont {B.}~\bibnamefont
  {{Chen}}}, \bibinfo {author} {\bibfnamefont {C.}~\bibnamefont {{Shen}}},
  \bibinfo {author} {\bibfnamefont {D.~E.}\ \bibnamefont {{Gary}}}, \bibinfo
  {author} {\bibfnamefont {K.~K.}\ \bibnamefont {{Reeves}}}, \bibinfo {author}
  {\bibfnamefont {G.~D.}\ \bibnamefont {{Fleishman}}}, \bibinfo {author}
  {\bibfnamefont {S.}~\bibnamefont {{Yu}}}, \bibinfo {author} {\bibfnamefont
  {F.}~\bibnamefont {{Guo}}}, \bibinfo {author} {\bibfnamefont
  {S.}~\bibnamefont {{Krucker}}}, \bibinfo {author} {\bibfnamefont
  {J.}~\bibnamefont {{Lin}}}, \bibinfo {author} {\bibfnamefont {G.~M.}\
  \bibnamefont {{Nita}}}, \ and\ \bibinfo {author} {\bibfnamefont
  {X.}~\bibnamefont {{Kong}}},\ }\href@noop {} {\bibfield  {journal} {\bibinfo
  {journal} {Nature Astronomy}\ }\textbf {\bibinfo {volume} {4}},\ \bibinfo
  {pages} {1140} (\bibinfo {year} {2020})}\BibitemShut {NoStop}%
\bibitem [{\citenamefont {{Drenkhahn}}\ and\ \citenamefont
  {{Spruit}}(2002)}]{drenkhahn02}%
  \BibitemOpen
  \bibfield  {author} {\bibinfo {author} {\bibfnamefont {G.}~\bibnamefont
  {{Drenkhahn}}}\ and\ \bibinfo {author} {\bibfnamefont {H.~C.}\ \bibnamefont
  {{Spruit}}},\ }\href@noop {} {\bibfield  {journal} {\bibinfo  {journal}
  {Astro. Astrophys.}\ }\textbf {\bibinfo {volume} {391}},\ \bibinfo {pages}
  {1141} (\bibinfo {year} {2002})}\BibitemShut {NoStop}%
\bibitem [{\citenamefont {{Sironi}}\ and\ \citenamefont
  {{Spitkovsky}}(2014)}]{sironi14}%
  \BibitemOpen
  \bibfield  {author} {\bibinfo {author} {\bibfnamefont {L.}~\bibnamefont
  {{Sironi}}}\ and\ \bibinfo {author} {\bibfnamefont {A.}~\bibnamefont
  {{Spitkovsky}}},\ }\href@noop {} {\bibfield  {journal} {\bibinfo  {journal}
  {Astrophys. J. Lett.}\ }\textbf {\bibinfo {volume} {783}},\ \bibinfo {eid}
  {L21} (\bibinfo {year} {2014})}\BibitemShut {NoStop}%
\bibitem [{\citenamefont {{Werner}}\ \emph {et~al.}(2016)\citenamefont
  {{Werner}}, \citenamefont {{Uzdensky}}, \citenamefont {{Cerutti}},
  \citenamefont {{Nalewajko}},\ and\ \citenamefont {{Begelman}}}]{werner16}%
  \BibitemOpen
  \bibfield  {author} {\bibinfo {author} {\bibfnamefont {G.~R.}\ \bibnamefont
  {{Werner}}}, \bibinfo {author} {\bibfnamefont {D.~A.}\ \bibnamefont
  {{Uzdensky}}}, \bibinfo {author} {\bibfnamefont {B.}~\bibnamefont
  {{Cerutti}}}, \bibinfo {author} {\bibfnamefont {K.}~\bibnamefont
  {{Nalewajko}}}, \ and\ \bibinfo {author} {\bibfnamefont {M.~C.}\ \bibnamefont
  {{Begelman}}},\ }\href@noop {} {\bibfield  {journal} {\bibinfo  {journal}
  {Astrophys. J. Lett.}\ }\textbf {\bibinfo {volume} {816}},\ \bibinfo {eid}
  {L8} (\bibinfo {year} {2016})}\BibitemShut {NoStop}%
\bibitem [{\citenamefont {{Li}}\ \emph {et~al.}(2017)\citenamefont {{Li}},
  \citenamefont {{Guo}}, \citenamefont {{Li}},\ and\ \citenamefont
  {{Li}}}]{li17}%
  \BibitemOpen
  \bibfield  {author} {\bibinfo {author} {\bibfnamefont {X.}~\bibnamefont
  {{Li}}}, \bibinfo {author} {\bibfnamefont {F.}~\bibnamefont {{Guo}}},
  \bibinfo {author} {\bibfnamefont {H.}~\bibnamefont {{Li}}}, \ and\ \bibinfo
  {author} {\bibfnamefont {G.}~\bibnamefont {{Li}}},\ }\href@noop {} {\bibfield
   {journal} {\bibinfo  {journal} {Astrophys. J.}\ }\textbf {\bibinfo {volume}
  {843}},\ \bibinfo {eid} {21} (\bibinfo {year} {2017})}\BibitemShut {NoStop}%
\bibitem [{\citenamefont {Dahlin}(2020)}]{dahlin20}%
  \BibitemOpen
  \bibfield  {author} {\bibinfo {author} {\bibfnamefont {J.~T.}\ \bibnamefont
  {Dahlin}},\ }\href@noop {} {\bibfield  {journal} {\bibinfo  {journal} {Phys.
  Plasmas}\ }\textbf {\bibinfo {volume} {27}},\ \bibinfo {pages} {100601}
  (\bibinfo {year} {2020})}\BibitemShut {NoStop}%
\bibitem [{\citenamefont {{Drake}}\ \emph {et~al.}(2006)\citenamefont
  {{Drake}}, \citenamefont {{Swisdak}}, \citenamefont {{Che}},\ and\
  \citenamefont {{Shay}}}]{drake06}%
  \BibitemOpen
  \bibfield  {author} {\bibinfo {author} {\bibfnamefont {J.~F.}\ \bibnamefont
  {{Drake}}}, \bibinfo {author} {\bibfnamefont {M.}~\bibnamefont {{Swisdak}}},
  \bibinfo {author} {\bibfnamefont {H.}~\bibnamefont {{Che}}}, \ and\ \bibinfo
  {author} {\bibfnamefont {M.~A.}\ \bibnamefont {{Shay}}},\ }\href@noop {}
  {\bibfield  {journal} {\bibinfo  {journal} {Nature}\ }\textbf {\bibinfo
  {volume} {443}},\ \bibinfo {pages} {553} (\bibinfo {year}
  {2006})}\BibitemShut {NoStop}%
\bibitem [{\citenamefont {{Hoshino}}\ \emph {et~al.}(2001)\citenamefont
  {{Hoshino}}, \citenamefont {{Mukai}}, \citenamefont {{Terasawa}},\ and\
  \citenamefont {{Shinohara}}}]{hoshino01}%
  \BibitemOpen
  \bibfield  {author} {\bibinfo {author} {\bibfnamefont {M.}~\bibnamefont
  {{Hoshino}}}, \bibinfo {author} {\bibfnamefont {T.}~\bibnamefont {{Mukai}}},
  \bibinfo {author} {\bibfnamefont {T.}~\bibnamefont {{Terasawa}}}, \ and\
  \bibinfo {author} {\bibfnamefont {I.}~\bibnamefont {{Shinohara}}},\
  }\href@noop {} {\bibfield  {journal} {\bibinfo  {journal} {J. Geophys. Res.}\
  }\textbf {\bibinfo {volume} {106}},\ \bibinfo {pages} {25979} (\bibinfo
  {year} {2001})}\BibitemShut {NoStop}%
\bibitem [{\citenamefont {Egedal}, \citenamefont {Le},\ and\ \citenamefont
  {Daughton}(2013)}]{egedal13}%
  \BibitemOpen
  \bibfield  {author} {\bibinfo {author} {\bibfnamefont {J.}~\bibnamefont
  {Egedal}}, \bibinfo {author} {\bibfnamefont {A.}~\bibnamefont {Le}}, \ and\
  \bibinfo {author} {\bibfnamefont {W.}~\bibnamefont {Daughton}},\ }\href@noop
  {} {\bibfield  {journal} {\bibinfo  {journal} {Phys. Plasmas}\ }\textbf
  {\bibinfo {volume} {20}},\ \bibinfo {pages} {061201} (\bibinfo {year}
  {2013})}\BibitemShut {NoStop}%
\bibitem [{\citenamefont {{Zenitani}}\ and\ \citenamefont
  {{Hoshino}}(2001)}]{zenitani01}%
  \BibitemOpen
  \bibfield  {author} {\bibinfo {author} {\bibfnamefont {S.}~\bibnamefont
  {{Zenitani}}}\ and\ \bibinfo {author} {\bibfnamefont {M.}~\bibnamefont
  {{Hoshino}}},\ }\href@noop {} {\bibfield  {journal} {\bibinfo  {journal}
  {Astrophys. J. Lett.}\ }\textbf {\bibinfo {volume} {562}},\ \bibinfo {pages}
  {L63} (\bibinfo {year} {2001})}\BibitemShut {NoStop}%
\bibitem [{\citenamefont {Gao}\ \emph {et~al.}(2016)\citenamefont {Gao},
  \citenamefont {Ji}, \citenamefont {Fiksel}, \citenamefont {Fox},
  \citenamefont {Evans},\ and\ \citenamefont {Alfonso}}]{Gao2016}%
  \BibitemOpen
  \bibfield  {author} {\bibinfo {author} {\bibfnamefont {L.}~\bibnamefont
  {Gao}}, \bibinfo {author} {\bibfnamefont {H.}~\bibnamefont {Ji}}, \bibinfo
  {author} {\bibfnamefont {G.}~\bibnamefont {Fiksel}}, \bibinfo {author}
  {\bibfnamefont {W.}~\bibnamefont {Fox}}, \bibinfo {author} {\bibfnamefont
  {M.}~\bibnamefont {Evans}}, \ and\ \bibinfo {author} {\bibfnamefont
  {N.}~\bibnamefont {Alfonso}},\ }\href@noop {} {\bibfield  {journal} {\bibinfo
   {journal} {Physics of Plasmas}\ }\textbf {\bibinfo {volume} {23}},\ \bibinfo
  {eid} {043106} (\bibinfo {year} {2016})}\BibitemShut {NoStop}%
\bibitem [{\citenamefont {Chien}\ \emph {et~al.}(2019)\citenamefont {Chien},
  \citenamefont {Gao}, \citenamefont {Ji}, \citenamefont {Xiaoxia},
  \citenamefont {Blackman}, \citenamefont {Chen}, \citenamefont {Efthimion},
  \citenamefont {Fiksel}, \citenamefont {Froula}, \citenamefont {Hill},
  \citenamefont {Huang}, \citenamefont {Lu}, \citenamefont {Moody},\ and\
  \citenamefont {Nilson}}]{chien2019}%
  \BibitemOpen
  \bibfield  {author} {\bibinfo {author} {\bibfnamefont {A.}~\bibnamefont
  {Chien}}, \bibinfo {author} {\bibfnamefont {L.}~\bibnamefont {Gao}}, \bibinfo
  {author} {\bibfnamefont {H.}~\bibnamefont {Ji}}, \bibinfo {author}
  {\bibfnamefont {Y.}~\bibnamefont {Xiaoxia}}, \bibinfo {author} {\bibfnamefont
  {E.}~\bibnamefont {Blackman}}, \bibinfo {author} {\bibfnamefont
  {H.}~\bibnamefont {Chen}}, \bibinfo {author} {\bibfnamefont {P.}~\bibnamefont
  {Efthimion}}, \bibinfo {author} {\bibfnamefont {G.}~\bibnamefont {Fiksel}},
  \bibinfo {author} {\bibfnamefont {D.}~\bibnamefont {Froula}}, \bibinfo
  {author} {\bibfnamefont {K.}~\bibnamefont {Hill}}, \bibinfo {author}
  {\bibfnamefont {K.}~\bibnamefont {Huang}}, \bibinfo {author} {\bibfnamefont
  {Q.}~\bibnamefont {Lu}}, \bibinfo {author} {\bibfnamefont {J.}~\bibnamefont
  {Moody}}, \ and\ \bibinfo {author} {\bibfnamefont {P.}~\bibnamefont
  {Nilson}},\ }\href {\doibase 10.1063/1.5095960} {\bibfield  {journal}
  {\bibinfo  {journal} {Physics of Plasmas}\ }\textbf {\bibinfo {volume}
  {26}},\ \bibinfo {pages} {062113} (\bibinfo {year} {2019})}\BibitemShut
  {NoStop}%
\bibitem [{\citenamefont {{Chien}}\ \emph {et~al.}(2021)\citenamefont
  {{Chien}}, \citenamefont {{Gao}}, \citenamefont {{Zhang}}, \citenamefont
  {{Ji}}, \citenamefont {{Blackman}}, \citenamefont {{Chen}}, \citenamefont
  {{Fiksel}}, \citenamefont {{Hill}},\ and\ \citenamefont
  {{Nilson}}}]{chien21}%
  \BibitemOpen
  \bibfield  {author} {\bibinfo {author} {\bibfnamefont {A.}~\bibnamefont
  {{Chien}}}, \bibinfo {author} {\bibfnamefont {L.}~\bibnamefont {{Gao}}},
  \bibinfo {author} {\bibfnamefont {S.}~\bibnamefont {{Zhang}}}, \bibinfo
  {author} {\bibfnamefont {H.}~\bibnamefont {{Ji}}}, \bibinfo {author}
  {\bibfnamefont {E.}~\bibnamefont {{Blackman}}}, \bibinfo {author}
  {\bibfnamefont {H.}~\bibnamefont {{Chen}}}, \bibinfo {author} {\bibfnamefont
  {G.}~\bibnamefont {{Fiksel}}}, \bibinfo {author} {\bibfnamefont
  {K.}~\bibnamefont {{Hill}}}, \ and\ \bibinfo {author} {\bibfnamefont
  {P.}~\bibnamefont {{Nilson}}},\ }\href@noop {} {\bibfield  {journal}
  {\bibinfo  {journal} {Phys. Plasmas}\ }\textbf {\bibinfo {volume} {28}},\
  \bibinfo {eid} {052105} (\bibinfo {year} {2021})}\BibitemShut {NoStop}%
\bibitem [{\citenamefont {Uzdensky}(2011)}]{uzdensky11}%
  \BibitemOpen
  \bibfield  {author} {\bibinfo {author} {\bibfnamefont {D.}~\bibnamefont
  {Uzdensky}},\ }\href@noop {} {\enquote {\bibinfo {title} {Magnetic
  reconnection in extreme astrophysical environments},}\ } (\bibinfo {year}
  {2011}),\ \bibinfo {note} {published in Space Sci. Rev., DOI:
  10.1007/s11214-011-9744-5}\BibitemShut {NoStop}%
\bibitem [{\citenamefont {Cerutti}\ \emph {et~al.}(2013)\citenamefont
  {Cerutti}, \citenamefont {Werner}, \citenamefont {Uzdensky},\ and\
  \citenamefont {Begelman}}]{cerutti13}%
  \BibitemOpen
  \bibfield  {author} {\bibinfo {author} {\bibfnamefont {B.}~\bibnamefont
  {Cerutti}}, \bibinfo {author} {\bibfnamefont {G.~R.}\ \bibnamefont {Werner}},
  \bibinfo {author} {\bibfnamefont {D.~A.}\ \bibnamefont {Uzdensky}}, \ and\
  \bibinfo {author} {\bibfnamefont {M.~C.}\ \bibnamefont {Begelman}},\
  }\href@noop {} {\bibfield  {journal} {\bibinfo  {journal} {Astrophys. J.}\
  }\textbf {\bibinfo {volume} {770}},\ \bibinfo {pages} {147} (\bibinfo {year}
  {2013})}\BibitemShut {NoStop}%
\bibitem [{\citenamefont {{Kroon}}\ \emph {et~al.}(2016)\citenamefont
  {{Kroon}}, \citenamefont {{Becker}}, \citenamefont {{Finke}},\ and\
  \citenamefont {{Dermer}}}]{kroon16}%
  \BibitemOpen
  \bibfield  {author} {\bibinfo {author} {\bibfnamefont {J.~J.}\ \bibnamefont
  {{Kroon}}}, \bibinfo {author} {\bibfnamefont {P.~A.}\ \bibnamefont
  {{Becker}}}, \bibinfo {author} {\bibfnamefont {J.~D.}\ \bibnamefont
  {{Finke}}}, \ and\ \bibinfo {author} {\bibfnamefont {C.~D.}\ \bibnamefont
  {{Dermer}}},\ }\href@noop {} {\bibfield  {journal} {\bibinfo  {journal}
  {Astrophys. J.}\ }\textbf {\bibinfo {volume} {833}},\ \bibinfo {eid} {157}
  (\bibinfo {year} {2016})}\BibitemShut {NoStop}%
\bibitem [{\citenamefont {Dahlin}, \citenamefont {Drake},\ and\ \citenamefont
  {Swisdak}(2014)}]{dahlin2014}%
  \BibitemOpen
  \bibfield  {author} {\bibinfo {author} {\bibfnamefont {J.~T.}\ \bibnamefont
  {Dahlin}}, \bibinfo {author} {\bibfnamefont {J.~F.}\ \bibnamefont {Drake}}, \
  and\ \bibinfo {author} {\bibfnamefont {M.}~\bibnamefont {Swisdak}},\ }\href
  {\doibase 10.1063/1.4894484} {\bibfield  {journal} {\bibinfo  {journal}
  {Physics of Plasmas}\ }\textbf {\bibinfo {volume} {21}},\ \bibinfo {pages}
  {092304} (\bibinfo {year} {2014})},\ \Eprint
  {http://arxiv.org/abs/https://doi.org/10.1063/1.4894484}
  {https://doi.org/10.1063/1.4894484} \BibitemShut {NoStop}%
\bibitem [{\citenamefont {Dahlin}, \citenamefont {Drake},\ and\ \citenamefont
  {Swisdak}(2016)}]{dahlin2016}%
  \BibitemOpen
  \bibfield  {author} {\bibinfo {author} {\bibfnamefont {J.~T.}\ \bibnamefont
  {Dahlin}}, \bibinfo {author} {\bibfnamefont {J.~F.}\ \bibnamefont {Drake}}, \
  and\ \bibinfo {author} {\bibfnamefont {M.}~\bibnamefont {Swisdak}},\ }\href
  {\doibase 10.1063/1.4972082} {\bibfield  {journal} {\bibinfo  {journal}
  {Physics of Plasmas}\ }\textbf {\bibinfo {volume} {23}},\ \bibinfo {pages}
  {120704} (\bibinfo {year} {2016})},\ \Eprint
  {http://arxiv.org/abs/https://doi.org/10.1063/1.4972082}
  {https://doi.org/10.1063/1.4972082} \BibitemShut {NoStop}%
\bibitem [{\citenamefont {Totorica}, \citenamefont {Abel},\ and\ \citenamefont
  {Fiuza}(2016)}]{totorica2016}%
  \BibitemOpen
  \bibfield  {author} {\bibinfo {author} {\bibfnamefont {S.~R.}\ \bibnamefont
  {Totorica}}, \bibinfo {author} {\bibfnamefont {T.}~\bibnamefont {Abel}}, \
  and\ \bibinfo {author} {\bibfnamefont {F.}~\bibnamefont {Fiuza}},\ }\href
  {\doibase 10.1103/PhysRevLett.116.095003} {\bibfield  {journal} {\bibinfo
  {journal} {Phys. Rev. Lett.}\ }\textbf {\bibinfo {volume} {116}},\ \bibinfo
  {pages} {095003} (\bibinfo {year} {2016})}\BibitemShut {NoStop}%
\bibitem [{\citenamefont {Savrukhin}(2001)}]{savrukhin01}%
  \BibitemOpen
  \bibfield  {author} {\bibinfo {author} {\bibfnamefont {P.~V.}\ \bibnamefont
  {Savrukhin}},\ }\href@noop {} {\bibfield  {journal} {\bibinfo  {journal}
  {Physical Review Letters}\ }\textbf {\bibinfo {volume} {86}},\ \bibinfo
  {pages} {3036} (\bibinfo {year} {2001})}\BibitemShut {NoStop}%
\bibitem [{\citenamefont {Klimanov}\ \emph {et~al.}(2007)\citenamefont
  {Klimanov}, \citenamefont {Fasoli}, \citenamefont {Goodman},\ and\
  \citenamefont {the TCV~team}}]{klimanov07}%
  \BibitemOpen
  \bibfield  {author} {\bibinfo {author} {\bibfnamefont {I.}~\bibnamefont
  {Klimanov}}, \bibinfo {author} {\bibfnamefont {A.}~\bibnamefont {Fasoli}},
  \bibinfo {author} {\bibfnamefont {T.~P.}\ \bibnamefont {Goodman}}, \ and\
  \bibinfo {author} {\bibnamefont {the TCV~team}},\ }\href@noop {} {\bibfield
  {journal} {\bibinfo  {journal} {Plasma Physics and Controlled Fusion}\
  }\textbf {\bibinfo {volume} {49}},\ \bibinfo {pages} {L1} (\bibinfo {year}
  {2007})}\BibitemShut {NoStop}%
\bibitem [{\citenamefont {DuBois}\ \emph {et~al.}(2017)\citenamefont {DuBois},
  \citenamefont {Almagri}, \citenamefont {Anderson}, \citenamefont
  {Den~Hartog}, \citenamefont {Lee},\ and\ \citenamefont {Sarff}}]{dubois2017}%
  \BibitemOpen
  \bibfield  {author} {\bibinfo {author} {\bibfnamefont {A.~M.}\ \bibnamefont
  {DuBois}}, \bibinfo {author} {\bibfnamefont {A.~F.}\ \bibnamefont {Almagri}},
  \bibinfo {author} {\bibfnamefont {J.~K.}\ \bibnamefont {Anderson}}, \bibinfo
  {author} {\bibfnamefont {D.~J.}\ \bibnamefont {Den~Hartog}}, \bibinfo
  {author} {\bibfnamefont {J.~D.}\ \bibnamefont {Lee}}, \ and\ \bibinfo
  {author} {\bibfnamefont {J.~S.}\ \bibnamefont {Sarff}},\ }\href {\doibase
  10.1103/PhysRevLett.118.075001} {\bibfield  {journal} {\bibinfo  {journal}
  {Phys. Rev. Lett.}\ }\textbf {\bibinfo {volume} {118}},\ \bibinfo {pages}
  {075001} (\bibinfo {year} {2017})}\BibitemShut {NoStop}%
\bibitem [{\citenamefont {{Nilson}}\ \emph {et~al.}(2006)\citenamefont
  {{Nilson}}, \citenamefont {{Willingale}}, \citenamefont {{Kaluza}},
  \citenamefont {{Kamperidis}}, \citenamefont {{Minardi}}, \citenamefont
  {{Wei}}, \citenamefont {{Fernandes}}, \citenamefont {{Notley}}, \citenamefont
  {{Bandyopadhyay}}, \citenamefont {{Sherlock}}, \citenamefont {{Kingham}},
  \citenamefont {{Tatarakis}}, \citenamefont {{Najmudin}}, \citenamefont
  {{Rozmus}}, \citenamefont {{Evans}}, \citenamefont {{Haines}}, \citenamefont
  {{Dangor}},\ and\ \citenamefont {{Krushelnick}}}]{nilson06}%
  \BibitemOpen
  \bibfield  {author} {\bibinfo {author} {\bibfnamefont {P.~M.}\ \bibnamefont
  {{Nilson}}}, \bibinfo {author} {\bibfnamefont {L.}~\bibnamefont
  {{Willingale}}}, \bibinfo {author} {\bibfnamefont {M.~C.}\ \bibnamefont
  {{Kaluza}}}, \bibinfo {author} {\bibfnamefont {C.}~\bibnamefont
  {{Kamperidis}}}, \bibinfo {author} {\bibfnamefont {S.}~\bibnamefont
  {{Minardi}}}, \bibinfo {author} {\bibfnamefont {M.~S.}\ \bibnamefont
  {{Wei}}}, \bibinfo {author} {\bibfnamefont {P.}~\bibnamefont {{Fernandes}}},
  \bibinfo {author} {\bibfnamefont {M.}~\bibnamefont {{Notley}}}, \bibinfo
  {author} {\bibfnamefont {S.}~\bibnamefont {{Bandyopadhyay}}}, \bibinfo
  {author} {\bibfnamefont {M.}~\bibnamefont {{Sherlock}}}, \bibinfo {author}
  {\bibfnamefont {R.~J.}\ \bibnamefont {{Kingham}}}, \bibinfo {author}
  {\bibfnamefont {M.}~\bibnamefont {{Tatarakis}}}, \bibinfo {author}
  {\bibfnamefont {Z.}~\bibnamefont {{Najmudin}}}, \bibinfo {author}
  {\bibfnamefont {W.}~\bibnamefont {{Rozmus}}}, \bibinfo {author}
  {\bibfnamefont {R.~G.}\ \bibnamefont {{Evans}}}, \bibinfo {author}
  {\bibfnamefont {M.~G.}\ \bibnamefont {{Haines}}}, \bibinfo {author}
  {\bibfnamefont {A.~E.}\ \bibnamefont {{Dangor}}}, \ and\ \bibinfo {author}
  {\bibfnamefont {K.}~\bibnamefont {{Krushelnick}}},\ }\href@noop {} {\bibfield
   {journal} {\bibinfo  {journal} {Phys. Rev. Lett.}\ }\textbf {\bibinfo
  {volume} {97}},\ \bibinfo {eid} {255001} (\bibinfo {year}
  {2006})}\BibitemShut {NoStop}%
\bibitem [{\citenamefont {{Willingale}}\ \emph {et~al.}(2010)\citenamefont
  {{Willingale}}, \citenamefont {{Nilson}}, \citenamefont {{Kaluza}},
  \citenamefont {{Dangor}}, \citenamefont {{Evans}}, \citenamefont
  {{Fernandes}}, \citenamefont {{Haines}}, \citenamefont {{Kamperidis}},
  \citenamefont {{Kingham}}, \citenamefont {{Ridgers}}, \citenamefont
  {{Sherlock}}, \citenamefont {{Thomas}}, \citenamefont {{Wei}}, \citenamefont
  {{Najmudin}}, \citenamefont {{Krushelnick}}, \citenamefont {{Bandyopadhyay}},
  \citenamefont {{Notley}}, \citenamefont {{Minardi}}, \citenamefont
  {{Tatarakis}},\ and\ \citenamefont {{Rozmus}}}]{willingale10}%
  \BibitemOpen
  \bibfield  {author} {\bibinfo {author} {\bibfnamefont {L.}~\bibnamefont
  {{Willingale}}}, \bibinfo {author} {\bibfnamefont {P.~M.}\ \bibnamefont
  {{Nilson}}}, \bibinfo {author} {\bibfnamefont {M.~C.}\ \bibnamefont
  {{Kaluza}}}, \bibinfo {author} {\bibfnamefont {A.~E.}\ \bibnamefont
  {{Dangor}}}, \bibinfo {author} {\bibfnamefont {R.~G.}\ \bibnamefont
  {{Evans}}}, \bibinfo {author} {\bibfnamefont {P.}~\bibnamefont
  {{Fernandes}}}, \bibinfo {author} {\bibfnamefont {M.~G.}\ \bibnamefont
  {{Haines}}}, \bibinfo {author} {\bibfnamefont {C.}~\bibnamefont
  {{Kamperidis}}}, \bibinfo {author} {\bibfnamefont {R.~J.}\ \bibnamefont
  {{Kingham}}}, \bibinfo {author} {\bibfnamefont {C.~P.}\ \bibnamefont
  {{Ridgers}}}, \bibinfo {author} {\bibfnamefont {M.}~\bibnamefont
  {{Sherlock}}}, \bibinfo {author} {\bibfnamefont {A.~G.~R.}\ \bibnamefont
  {{Thomas}}}, \bibinfo {author} {\bibfnamefont {M.~S.}\ \bibnamefont {{Wei}}},
  \bibinfo {author} {\bibfnamefont {Z.}~\bibnamefont {{Najmudin}}}, \bibinfo
  {author} {\bibfnamefont {K.}~\bibnamefont {{Krushelnick}}}, \bibinfo {author}
  {\bibfnamefont {S.}~\bibnamefont {{Bandyopadhyay}}}, \bibinfo {author}
  {\bibfnamefont {M.}~\bibnamefont {{Notley}}}, \bibinfo {author}
  {\bibfnamefont {S.}~\bibnamefont {{Minardi}}}, \bibinfo {author}
  {\bibfnamefont {M.}~\bibnamefont {{Tatarakis}}}, \ and\ \bibinfo {author}
  {\bibfnamefont {W.}~\bibnamefont {{Rozmus}}},\ }\href@noop {} {\bibfield
  {journal} {\bibinfo  {journal} {Phys. Plasmas}\ }\textbf {\bibinfo {volume}
  {17}},\ \bibinfo {pages} {043104} (\bibinfo {year} {2010})}\BibitemShut
  {NoStop}%
\bibitem [{\citenamefont {{Zhong}}\ \emph {et~al.}(2010)\citenamefont
  {{Zhong}}, \citenamefont {{Li}}, \citenamefont {{Wang}}, \citenamefont
  {{Wang}}, \citenamefont {{Dong}}, \citenamefont {{Xiao}}, \citenamefont
  {{Wang}}, \citenamefont {{Liu}}, \citenamefont {{Zhang}}, \citenamefont
  {{An}}, \citenamefont {{Wang}}, \citenamefont {{Zhu}}, \citenamefont {{Gu}},
  \citenamefont {{He}}, \citenamefont {{Zhao}},\ and\ \citenamefont
  {{Zhang}}}]{zhong10}%
  \BibitemOpen
  \bibfield  {author} {\bibinfo {author} {\bibfnamefont {J.}~\bibnamefont
  {{Zhong}}}, \bibinfo {author} {\bibfnamefont {Y.}~\bibnamefont {{Li}}},
  \bibinfo {author} {\bibfnamefont {X.}~\bibnamefont {{Wang}}}, \bibinfo
  {author} {\bibfnamefont {J.}~\bibnamefont {{Wang}}}, \bibinfo {author}
  {\bibfnamefont {Q.}~\bibnamefont {{Dong}}}, \bibinfo {author} {\bibfnamefont
  {C.}~\bibnamefont {{Xiao}}}, \bibinfo {author} {\bibfnamefont
  {S.}~\bibnamefont {{Wang}}}, \bibinfo {author} {\bibfnamefont
  {X.}~\bibnamefont {{Liu}}}, \bibinfo {author} {\bibfnamefont
  {L.}~\bibnamefont {{Zhang}}}, \bibinfo {author} {\bibfnamefont
  {L.}~\bibnamefont {{An}}}, \bibinfo {author} {\bibfnamefont {F.}~\bibnamefont
  {{Wang}}}, \bibinfo {author} {\bibfnamefont {J.}~\bibnamefont {{Zhu}}},
  \bibinfo {author} {\bibfnamefont {Y.}~\bibnamefont {{Gu}}}, \bibinfo {author}
  {\bibfnamefont {X.}~\bibnamefont {{He}}}, \bibinfo {author} {\bibfnamefont
  {G.}~\bibnamefont {{Zhao}}}, \ and\ \bibinfo {author} {\bibfnamefont
  {J.}~\bibnamefont {{Zhang}}},\ }\href@noop {} {\bibfield  {journal} {\bibinfo
   {journal} {Nature Phys.}\ }\textbf {\bibinfo {volume} {6}},\ \bibinfo
  {pages} {984} (\bibinfo {year} {2010})}\BibitemShut {NoStop}%
\bibitem [{\citenamefont {Fiksel}\ \emph {et~al.}(2014)\citenamefont {Fiksel},
  \citenamefont {Fox}, \citenamefont {Bhattacharjee}, \citenamefont {Barnak},
  \citenamefont {Chang}, \citenamefont {Germaschewski}, \citenamefont {Hu},\
  and\ \citenamefont {Nilson}}]{fiksel14}%
  \BibitemOpen
  \bibfield  {author} {\bibinfo {author} {\bibfnamefont {G.}~\bibnamefont
  {Fiksel}}, \bibinfo {author} {\bibfnamefont {W.}~\bibnamefont {Fox}},
  \bibinfo {author} {\bibfnamefont {A.}~\bibnamefont {Bhattacharjee}}, \bibinfo
  {author} {\bibfnamefont {D.~H.}\ \bibnamefont {Barnak}}, \bibinfo {author}
  {\bibfnamefont {P.~Y.}\ \bibnamefont {Chang}}, \bibinfo {author}
  {\bibfnamefont {K.}~\bibnamefont {Germaschewski}}, \bibinfo {author}
  {\bibfnamefont {S.~X.}\ \bibnamefont {Hu}}, \ and\ \bibinfo {author}
  {\bibfnamefont {P.~M.}\ \bibnamefont {Nilson}},\ }\href@noop {} {\bibfield
  {journal} {\bibinfo  {journal} {Phys. Rev. Lett.}\ }\textbf {\bibinfo
  {volume} {113}},\ \bibinfo {pages} {105003} (\bibinfo {year}
  {2014})}\BibitemShut {NoStop}%
\bibitem [{\citenamefont {E~Raymond}\ \emph {et~al.}(2018)\citenamefont
  {E~Raymond}, \citenamefont {Dong}, \citenamefont {Mckelvey}, \citenamefont
  {Zulick}, \citenamefont {Alexander}, \citenamefont {Bhattacharjee},
  \citenamefont {T~Campbell}, \citenamefont {Chen}, \citenamefont {Chvykov},
  \citenamefont {Del~Rio}, \citenamefont {Fitzsimmons}, \citenamefont {Fox},
  \citenamefont {Hou}, \citenamefont {Maksimchuk}, \citenamefont {Mileham},
  \citenamefont {Nees}, \citenamefont {M~Nilson}, \citenamefont {Stoeckl},
  \citenamefont {Thomas},\ and\ \citenamefont {Willingale}}]{raymond2018}%
  \BibitemOpen
  \bibfield  {author} {\bibinfo {author} {\bibfnamefont {A.}~\bibnamefont
  {E~Raymond}}, \bibinfo {author} {\bibfnamefont {C.}~\bibnamefont {Dong}},
  \bibinfo {author} {\bibfnamefont {A.}~\bibnamefont {Mckelvey}}, \bibinfo
  {author} {\bibfnamefont {C.}~\bibnamefont {Zulick}}, \bibinfo {author}
  {\bibfnamefont {N.}~\bibnamefont {Alexander}}, \bibinfo {author}
  {\bibfnamefont {A.}~\bibnamefont {Bhattacharjee}}, \bibinfo {author}
  {\bibfnamefont {P.}~\bibnamefont {T~Campbell}}, \bibinfo {author}
  {\bibfnamefont {H.}~\bibnamefont {Chen}}, \bibinfo {author} {\bibfnamefont
  {V.}~\bibnamefont {Chvykov}}, \bibinfo {author} {\bibfnamefont
  {E.}~\bibnamefont {Del~Rio}}, \bibinfo {author} {\bibfnamefont
  {P.}~\bibnamefont {Fitzsimmons}}, \bibinfo {author} {\bibfnamefont
  {W.}~\bibnamefont {Fox}}, \bibinfo {author} {\bibfnamefont {B.}~\bibnamefont
  {Hou}}, \bibinfo {author} {\bibfnamefont {A.}~\bibnamefont {Maksimchuk}},
  \bibinfo {author} {\bibfnamefont {C.}~\bibnamefont {Mileham}}, \bibinfo
  {author} {\bibfnamefont {J.}~\bibnamefont {Nees}}, \bibinfo {author}
  {\bibfnamefont {P.}~\bibnamefont {M~Nilson}}, \bibinfo {author}
  {\bibfnamefont {C.}~\bibnamefont {Stoeckl}}, \bibinfo {author} {\bibfnamefont
  {A.}~\bibnamefont {Thomas}}, \ and\ \bibinfo {author} {\bibfnamefont
  {L.}~\bibnamefont {Willingale}},\ }\href {\doibase
  10.1103/PhysRevE.98.043207} {\bibfield  {journal} {\bibinfo  {journal}
  {Physical Review E}\ }\textbf {\bibinfo {volume} {98}},\ \bibinfo {pages}
  {043207} (\bibinfo {year} {2018})}\BibitemShut {NoStop}%
\bibitem [{\citenamefont {Dong}\ \emph {et~al.}(2012)\citenamefont {Dong},
  \citenamefont {Wang}, \citenamefont {Lu}, \citenamefont {Huang},
  \citenamefont {Yuan}, \citenamefont {Liu}, \citenamefont {Lin}, \citenamefont
  {Li}, \citenamefont {Wei}, \citenamefont {Zhong}, \citenamefont {Shi},
  \citenamefont {Jiang}, \citenamefont {Ding}, \citenamefont {Jiang},
  \citenamefont {Du}, \citenamefont {He}, \citenamefont {Yu}, \citenamefont
  {Liu}, \citenamefont {Wang}, \citenamefont {Tang}, \citenamefont {Zhu},
  \citenamefont {Zhao}, \citenamefont {Sheng},\ and\ \citenamefont
  {Zhang}}]{dong12}%
  \BibitemOpen
  \bibfield  {author} {\bibinfo {author} {\bibfnamefont {Q.-L.}\ \bibnamefont
  {Dong}}, \bibinfo {author} {\bibfnamefont {S.-J.}\ \bibnamefont {Wang}},
  \bibinfo {author} {\bibfnamefont {Q.-M.}\ \bibnamefont {Lu}}, \bibinfo
  {author} {\bibfnamefont {C.}~\bibnamefont {Huang}}, \bibinfo {author}
  {\bibfnamefont {D.-W.}\ \bibnamefont {Yuan}}, \bibinfo {author}
  {\bibfnamefont {X.}~\bibnamefont {Liu}}, \bibinfo {author} {\bibfnamefont
  {X.-X.}\ \bibnamefont {Lin}}, \bibinfo {author} {\bibfnamefont {Y.-T.}\
  \bibnamefont {Li}}, \bibinfo {author} {\bibfnamefont {H.-G.}\ \bibnamefont
  {Wei}}, \bibinfo {author} {\bibfnamefont {J.-Y.}\ \bibnamefont {Zhong}},
  \bibinfo {author} {\bibfnamefont {J.-R.}\ \bibnamefont {Shi}}, \bibinfo
  {author} {\bibfnamefont {S.-E.}\ \bibnamefont {Jiang}}, \bibinfo {author}
  {\bibfnamefont {Y.-K.}\ \bibnamefont {Ding}}, \bibinfo {author}
  {\bibfnamefont {B.-B.}\ \bibnamefont {Jiang}}, \bibinfo {author}
  {\bibfnamefont {K.}~\bibnamefont {Du}}, \bibinfo {author} {\bibfnamefont
  {X.-T.}\ \bibnamefont {He}}, \bibinfo {author} {\bibfnamefont {M.~Y.}\
  \bibnamefont {Yu}}, \bibinfo {author} {\bibfnamefont {C.~S.}\ \bibnamefont
  {Liu}}, \bibinfo {author} {\bibfnamefont {S.}~\bibnamefont {Wang}}, \bibinfo
  {author} {\bibfnamefont {Y.-J.}\ \bibnamefont {Tang}}, \bibinfo {author}
  {\bibfnamefont {J.-Q.}\ \bibnamefont {Zhu}}, \bibinfo {author} {\bibfnamefont
  {G.}~\bibnamefont {Zhao}}, \bibinfo {author} {\bibfnamefont {Z.-M.}\
  \bibnamefont {Sheng}}, \ and\ \bibinfo {author} {\bibfnamefont
  {J.}~\bibnamefont {Zhang}},\ }\href@noop {} {\bibfield  {journal} {\bibinfo
  {journal} {Phys. Rev. Lett.}\ }\textbf {\bibinfo {volume} {108}},\ \bibinfo
  {pages} {215001} (\bibinfo {year} {2012})}\BibitemShut {NoStop}%
\bibitem [{\citenamefont {Zhong}\ \emph {et~al.}(2016)\citenamefont {Zhong},
  \citenamefont {Lin}, \citenamefont {Li}, \citenamefont {Wang}, \citenamefont
  {Li}, \citenamefont {Zhang}, \citenamefont {Yuan}, \citenamefont {Ping},
  \citenamefont {Wei}, \citenamefont {Wang}, \citenamefont {Su}, \citenamefont
  {Li}, \citenamefont {Han}, \citenamefont {Liao}, \citenamefont {Yin},
  \citenamefont {Fang}, \citenamefont {Yuan}, \citenamefont {Wang},
  \citenamefont {Sun}, \citenamefont {Liang}, \citenamefont {Wang},
  \citenamefont {Ding}, \citenamefont {He}, \citenamefont {Zhu}, \citenamefont
  {Sheng}, \citenamefont {Li}, \citenamefont {Zhao},\ and\ \citenamefont
  {Zhang}}]{Zhong_2016}%
  \BibitemOpen
  \bibfield  {author} {\bibinfo {author} {\bibfnamefont {J.~Y.}\ \bibnamefont
  {Zhong}}, \bibinfo {author} {\bibfnamefont {J.}~\bibnamefont {Lin}}, \bibinfo
  {author} {\bibfnamefont {Y.~T.}\ \bibnamefont {Li}}, \bibinfo {author}
  {\bibfnamefont {X.}~\bibnamefont {Wang}}, \bibinfo {author} {\bibfnamefont
  {Y.}~\bibnamefont {Li}}, \bibinfo {author} {\bibfnamefont {K.}~\bibnamefont
  {Zhang}}, \bibinfo {author} {\bibfnamefont {D.~W.}\ \bibnamefont {Yuan}},
  \bibinfo {author} {\bibfnamefont {Y.~L.}\ \bibnamefont {Ping}}, \bibinfo
  {author} {\bibfnamefont {H.~G.}\ \bibnamefont {Wei}}, \bibinfo {author}
  {\bibfnamefont {J.~Q.}\ \bibnamefont {Wang}}, \bibinfo {author}
  {\bibfnamefont {L.~N.}\ \bibnamefont {Su}}, \bibinfo {author} {\bibfnamefont
  {F.}~\bibnamefont {Li}}, \bibinfo {author} {\bibfnamefont {B.}~\bibnamefont
  {Han}}, \bibinfo {author} {\bibfnamefont {G.~Q.}\ \bibnamefont {Liao}},
  \bibinfo {author} {\bibfnamefont {C.~L.}\ \bibnamefont {Yin}}, \bibinfo
  {author} {\bibfnamefont {Y.}~\bibnamefont {Fang}}, \bibinfo {author}
  {\bibfnamefont {X.}~\bibnamefont {Yuan}}, \bibinfo {author} {\bibfnamefont
  {C.}~\bibnamefont {Wang}}, \bibinfo {author} {\bibfnamefont {J.~R.}\
  \bibnamefont {Sun}}, \bibinfo {author} {\bibfnamefont {G.~Y.}\ \bibnamefont
  {Liang}}, \bibinfo {author} {\bibfnamefont {F.~L.}\ \bibnamefont {Wang}},
  \bibinfo {author} {\bibfnamefont {Y.~K.}\ \bibnamefont {Ding}}, \bibinfo
  {author} {\bibfnamefont {X.~T.}\ \bibnamefont {He}}, \bibinfo {author}
  {\bibfnamefont {J.~Q.}\ \bibnamefont {Zhu}}, \bibinfo {author} {\bibfnamefont
  {Z.~M.}\ \bibnamefont {Sheng}}, \bibinfo {author} {\bibfnamefont
  {G.}~\bibnamefont {Li}}, \bibinfo {author} {\bibfnamefont {G.}~\bibnamefont
  {Zhao}}, \ and\ \bibinfo {author} {\bibfnamefont {J.}~\bibnamefont {Zhang}},\
  }\href {\doibase 10.3847/0067-0049/225/2/30} {\bibfield  {journal} {\bibinfo
  {journal} {The Astrophysical Journal Supplement Series}\ }\textbf {\bibinfo
  {volume} {225}},\ \bibinfo {pages} {30} (\bibinfo {year} {2016})}\BibitemShut
  {NoStop}%
\bibitem [{\citenamefont {Wilks}\ \emph {et~al.}(2001)\citenamefont {Wilks},
  \citenamefont {Langdon}, \citenamefont {Cowan}, \citenamefont {Roth},
  \citenamefont {Singh}, \citenamefont {Hatchett}, \citenamefont {Key},
  \citenamefont {Pennington}, \citenamefont {MacKinnon},\ and\ \citenamefont
  {Snavely}}]{wilks2001}%
  \BibitemOpen
  \bibfield  {author} {\bibinfo {author} {\bibfnamefont {S.~C.}\ \bibnamefont
  {Wilks}}, \bibinfo {author} {\bibfnamefont {A.~B.}\ \bibnamefont {Langdon}},
  \bibinfo {author} {\bibfnamefont {T.~E.}\ \bibnamefont {Cowan}}, \bibinfo
  {author} {\bibfnamefont {M.}~\bibnamefont {Roth}}, \bibinfo {author}
  {\bibfnamefont {M.}~\bibnamefont {Singh}}, \bibinfo {author} {\bibfnamefont
  {S.}~\bibnamefont {Hatchett}}, \bibinfo {author} {\bibfnamefont {M.~H.}\
  \bibnamefont {Key}}, \bibinfo {author} {\bibfnamefont {D.}~\bibnamefont
  {Pennington}}, \bibinfo {author} {\bibfnamefont {A.}~\bibnamefont
  {MacKinnon}}, \ and\ \bibinfo {author} {\bibfnamefont {R.~A.}\ \bibnamefont
  {Snavely}},\ }\href@noop {} {\bibfield  {journal} {\bibinfo  {journal}
  {Physics of Plasmas}\ }\textbf {\bibinfo {volume} {8}},\ \bibinfo {pages}
  {542} (\bibinfo {year} {2001})}\BibitemShut {NoStop}%
\bibitem [{\citenamefont {Yamada}\ \emph {et~al.}(1997)\citenamefont {Yamada},
  \citenamefont {Ji}, \citenamefont {Hsu}, \citenamefont {Carter},
  \citenamefont {Kulsrud}, \citenamefont {Bretz}, \citenamefont {Jobes},
  \citenamefont {Ono},\ and\ \citenamefont {Perkins}}]{YamadaPOP1997}%
  \BibitemOpen
  \bibfield  {author} {\bibinfo {author} {\bibfnamefont {M.}~\bibnamefont
  {Yamada}}, \bibinfo {author} {\bibfnamefont {H.}~\bibnamefont {Ji}}, \bibinfo
  {author} {\bibfnamefont {S.}~\bibnamefont {Hsu}}, \bibinfo {author}
  {\bibfnamefont {T.}~\bibnamefont {Carter}}, \bibinfo {author} {\bibfnamefont
  {R.}~\bibnamefont {Kulsrud}}, \bibinfo {author} {\bibfnamefont
  {N.}~\bibnamefont {Bretz}}, \bibinfo {author} {\bibfnamefont
  {F.}~\bibnamefont {Jobes}}, \bibinfo {author} {\bibfnamefont
  {Y.}~\bibnamefont {Ono}}, \ and\ \bibinfo {author} {\bibfnamefont
  {F.}~\bibnamefont {Perkins}},\ }\href@noop {} {\bibfield  {journal} {\bibinfo
   {journal} {Physics of Plasmas}\ }\textbf {\bibinfo {volume} {4}},\ \bibinfo
  {pages} {1936} (\bibinfo {year} {1997})}\BibitemShut {NoStop}%
\bibitem [{\citenamefont {{Phan}}\ \emph {et~al.}(2018)\citenamefont {{Phan}},
  \citenamefont {{Eastwood}}, \citenamefont {{Shay}}, \citenamefont {{Drake}},
  \citenamefont {{Sonnerup}}, \citenamefont {{Fujimoto}}, \citenamefont
  {{Cassak}}, \citenamefont {{{\O}ieroset}}, \citenamefont {{Burch}},
  \citenamefont {{Torbert}}, \citenamefont {{Rager}}, \citenamefont
  {{Dorelli}}, \citenamefont {{Gershman}}, \citenamefont {{Pollock}},
  \citenamefont {{Pyakurel}}, \citenamefont {{Haggerty}}, \citenamefont
  {{Khotyaintsev}}, \citenamefont {{Lavraud}}, \citenamefont {{Saito}},
  \citenamefont {{Oka}}, \citenamefont {{Ergun}}, \citenamefont {{Retino}},
  \citenamefont {{Le Contel}}, \citenamefont {{Argall}}, \citenamefont
  {{Giles}}, \citenamefont {{Moore}}, \citenamefont {{Wilder}}, \citenamefont
  {{Strangeway}}, \citenamefont {{Russell}}, \citenamefont {{Lindqvist}},\ and\
  \citenamefont {{Magnes}}}]{phan18}%
  \BibitemOpen
  \bibfield  {author} {\bibinfo {author} {\bibfnamefont {T.~D.}\ \bibnamefont
  {{Phan}}}, \bibinfo {author} {\bibfnamefont {J.~P.}\ \bibnamefont
  {{Eastwood}}}, \bibinfo {author} {\bibfnamefont {M.~A.}\ \bibnamefont
  {{Shay}}}, \bibinfo {author} {\bibfnamefont {J.~F.}\ \bibnamefont {{Drake}}},
  \bibinfo {author} {\bibfnamefont {B.~U.~{\"O}.}\ \bibnamefont {{Sonnerup}}},
  \bibinfo {author} {\bibfnamefont {M.}~\bibnamefont {{Fujimoto}}}, \bibinfo
  {author} {\bibfnamefont {P.~A.}\ \bibnamefont {{Cassak}}}, \bibinfo {author}
  {\bibfnamefont {M.}~\bibnamefont {{{\O}ieroset}}}, \bibinfo {author}
  {\bibfnamefont {J.~L.}\ \bibnamefont {{Burch}}}, \bibinfo {author}
  {\bibfnamefont {R.~B.}\ \bibnamefont {{Torbert}}}, \bibinfo {author}
  {\bibfnamefont {A.~C.}\ \bibnamefont {{Rager}}}, \bibinfo {author}
  {\bibfnamefont {J.~C.}\ \bibnamefont {{Dorelli}}}, \bibinfo {author}
  {\bibfnamefont {D.~J.}\ \bibnamefont {{Gershman}}}, \bibinfo {author}
  {\bibfnamefont {C.}~\bibnamefont {{Pollock}}}, \bibinfo {author}
  {\bibfnamefont {P.~S.}\ \bibnamefont {{Pyakurel}}}, \bibinfo {author}
  {\bibfnamefont {C.~C.}\ \bibnamefont {{Haggerty}}}, \bibinfo {author}
  {\bibfnamefont {Y.}~\bibnamefont {{Khotyaintsev}}}, \bibinfo {author}
  {\bibfnamefont {B.}~\bibnamefont {{Lavraud}}}, \bibinfo {author}
  {\bibfnamefont {Y.}~\bibnamefont {{Saito}}}, \bibinfo {author} {\bibfnamefont
  {M.}~\bibnamefont {{Oka}}}, \bibinfo {author} {\bibfnamefont {R.~E.}\
  \bibnamefont {{Ergun}}}, \bibinfo {author} {\bibfnamefont {A.}~\bibnamefont
  {{Retino}}}, \bibinfo {author} {\bibfnamefont {O.}~\bibnamefont {{Le
  Contel}}}, \bibinfo {author} {\bibfnamefont {M.~R.}\ \bibnamefont
  {{Argall}}}, \bibinfo {author} {\bibfnamefont {B.~L.}\ \bibnamefont
  {{Giles}}}, \bibinfo {author} {\bibfnamefont {T.~E.}\ \bibnamefont
  {{Moore}}}, \bibinfo {author} {\bibfnamefont {F.~D.}\ \bibnamefont
  {{Wilder}}}, \bibinfo {author} {\bibfnamefont {R.~J.}\ \bibnamefont
  {{Strangeway}}}, \bibinfo {author} {\bibfnamefont {C.~T.}\ \bibnamefont
  {{Russell}}}, \bibinfo {author} {\bibfnamefont {P.~A.}\ \bibnamefont
  {{Lindqvist}}}, \ and\ \bibinfo {author} {\bibfnamefont {W.}~\bibnamefont
  {{Magnes}}},\ }\href@noop {} {\bibfield  {journal} {\bibinfo  {journal}
  {Nature}\ }\textbf {\bibinfo {volume} {557}},\ \bibinfo {pages} {202}
  (\bibinfo {year} {2018})}\BibitemShut {NoStop}%
\bibitem [{\citenamefont {Drake}\ \emph {et~al.}(1984)\citenamefont {Drake},
  \citenamefont {Turner}, \citenamefont {Lasinski}, \citenamefont {Estabrook},
  \citenamefont {Campbell}, \citenamefont {Wang}, \citenamefont {Phillion},
  \citenamefont {Williams},\ and\ \citenamefont {Kruer}}]{drake_1984}%
  \BibitemOpen
  \bibfield  {author} {\bibinfo {author} {\bibfnamefont {R.~P.}\ \bibnamefont
  {Drake}}, \bibinfo {author} {\bibfnamefont {R.~E.}\ \bibnamefont {Turner}},
  \bibinfo {author} {\bibfnamefont {B.~F.}\ \bibnamefont {Lasinski}}, \bibinfo
  {author} {\bibfnamefont {K.~G.}\ \bibnamefont {Estabrook}}, \bibinfo {author}
  {\bibfnamefont {E.~M.}\ \bibnamefont {Campbell}}, \bibinfo {author}
  {\bibfnamefont {C.~L.}\ \bibnamefont {Wang}}, \bibinfo {author}
  {\bibfnamefont {D.~W.}\ \bibnamefont {Phillion}}, \bibinfo {author}
  {\bibfnamefont {E.~A.}\ \bibnamefont {Williams}}, \ and\ \bibinfo {author}
  {\bibfnamefont {W.~L.}\ \bibnamefont {Kruer}},\ }\href {\doibase
  10.1103/PhysRevLett.53.1739} {\bibfield  {journal} {\bibinfo  {journal}
  {Phys. Rev. Lett.}\ }\textbf {\bibinfo {volume} {53}},\ \bibinfo {pages}
  {1739} (\bibinfo {year} {1984})}\BibitemShut {NoStop}%
\bibitem [{\citenamefont {Figueroa}\ \emph {et~al.}(1984)\citenamefont
  {Figueroa}, \citenamefont {Joshi}, \citenamefont {Azechi}, \citenamefont
  {Ebrahim},\ and\ \citenamefont {Estabrook}}]{figueroa_1984}%
  \BibitemOpen
  \bibfield  {author} {\bibinfo {author} {\bibfnamefont {H.}~\bibnamefont
  {Figueroa}}, \bibinfo {author} {\bibfnamefont {C.}~\bibnamefont {Joshi}},
  \bibinfo {author} {\bibfnamefont {H.}~\bibnamefont {Azechi}}, \bibinfo
  {author} {\bibfnamefont {N.~A.}\ \bibnamefont {Ebrahim}}, \ and\ \bibinfo
  {author} {\bibfnamefont {K.}~\bibnamefont {Estabrook}},\ }\href {\doibase
  10.1063/1.864801} {\bibfield  {journal} {\bibinfo  {journal} {The Physics of
  Fluids}\ }\textbf {\bibinfo {volume} {27}},\ \bibinfo {pages} {1887}
  (\bibinfo {year} {1984})},\ \Eprint
  {http://arxiv.org/abs/https://aip.scitation.org/doi/pdf/10.1063/1.864801}
  {https://aip.scitation.org/doi/pdf/10.1063/1.864801} \BibitemShut {NoStop}%
\bibitem [{\citenamefont {Ebrahim}\ \emph {et~al.}(1980)\citenamefont
  {Ebrahim}, \citenamefont {Baldis}, \citenamefont {Joshi},\ and\ \citenamefont
  {Benesch}}]{ebrahim_1980}%
  \BibitemOpen
  \bibfield  {author} {\bibinfo {author} {\bibfnamefont {N.~A.}\ \bibnamefont
  {Ebrahim}}, \bibinfo {author} {\bibfnamefont {H.~A.}\ \bibnamefont {Baldis}},
  \bibinfo {author} {\bibfnamefont {C.}~\bibnamefont {Joshi}}, \ and\ \bibinfo
  {author} {\bibfnamefont {R.}~\bibnamefont {Benesch}},\ }\href {\doibase
  10.1103/PhysRevLett.45.1179} {\bibfield  {journal} {\bibinfo  {journal}
  {Phys. Rev. Lett.}\ }\textbf {\bibinfo {volume} {45}},\ \bibinfo {pages}
  {1179} (\bibinfo {year} {1980})}\BibitemShut {NoStop}%
\bibitem [{\citenamefont {Bowers}\ \emph {et~al.}(2008)\citenamefont {Bowers},
  \citenamefont {Albright}, \citenamefont {Yin}, \citenamefont {Bergen},\ and\
  \citenamefont {Kwan}}]{bowers08}%
  \BibitemOpen
  \bibfield  {author} {\bibinfo {author} {\bibfnamefont {K.~J.}\ \bibnamefont
  {Bowers}}, \bibinfo {author} {\bibfnamefont {B.~J.}\ \bibnamefont
  {Albright}}, \bibinfo {author} {\bibfnamefont {L.}~\bibnamefont {Yin}},
  \bibinfo {author} {\bibfnamefont {B.}~\bibnamefont {Bergen}}, \ and\ \bibinfo
  {author} {\bibfnamefont {T.~J.~T.}\ \bibnamefont {Kwan}},\ }\href@noop {}
  {\bibfield  {journal} {\bibinfo  {journal} {Phys. Plasmas}\ }\textbf
  {\bibinfo {volume} {15}},\ \bibinfo {pages} {055703} (\bibinfo {year}
  {2008})}\BibitemShut {NoStop}%
\bibitem [{\citenamefont {Uzdensky}\ and\ \citenamefont
  {Kulsrud}(2006)}]{uzdensky06}%
  \BibitemOpen
  \bibfield  {author} {\bibinfo {author} {\bibfnamefont {D.}~\bibnamefont
  {Uzdensky}}\ and\ \bibinfo {author} {\bibfnamefont {R.}~\bibnamefont
  {Kulsrud}},\ }\href@noop {} {\bibfield  {journal} {\bibinfo  {journal} {Phys.
  Plasmas}\ }\textbf {\bibinfo {volume} {13}},\ \bibinfo {pages} {062305}
  (\bibinfo {year} {2006})}\BibitemShut {NoStop}%
\bibitem [{\citenamefont {Birn}\ \emph {et~al.}(2001)\citenamefont {Birn},
  \citenamefont {Drake}, \citenamefont {Shay}, \citenamefont {Rogers},
  \citenamefont {Denton}, \citenamefont {Hesse}, \citenamefont {Kuznetsova},
  \citenamefont {Ma}, \citenamefont {Bhattachargee}, \citenamefont {Otto},\
  and\ \citenamefont {Pritchett}}]{birn01}%
  \BibitemOpen
  \bibfield  {author} {\bibinfo {author} {\bibfnamefont {J.}~\bibnamefont
  {Birn}}, \bibinfo {author} {\bibfnamefont {J.}~\bibnamefont {Drake}},
  \bibinfo {author} {\bibfnamefont {M.}~\bibnamefont {Shay}}, \bibinfo {author}
  {\bibfnamefont {B.}~\bibnamefont {Rogers}}, \bibinfo {author} {\bibfnamefont
  {R.}~\bibnamefont {Denton}}, \bibinfo {author} {\bibfnamefont
  {M.}~\bibnamefont {Hesse}}, \bibinfo {author} {\bibfnamefont
  {M.}~\bibnamefont {Kuznetsova}}, \bibinfo {author} {\bibfnamefont
  {Z.}~\bibnamefont {Ma}}, \bibinfo {author} {\bibfnamefont {A.}~\bibnamefont
  {Bhattachargee}}, \bibinfo {author} {\bibfnamefont {A.}~\bibnamefont {Otto}},
  \ and\ \bibinfo {author} {\bibfnamefont {P.}~\bibnamefont {Pritchett}},\
  }\href@noop {} {\bibfield  {journal} {\bibinfo  {journal} {J. Geophys. Res.}\
  }\textbf {\bibinfo {volume} {106}},\ \bibinfo {pages} {3715} (\bibinfo {year}
  {2001})}\BibitemShut {NoStop}%
\bibitem [{\citenamefont {Cassak}, \citenamefont {Liu},\ and\ \citenamefont
  {Shay}(2017)}]{cassak17}%
  \BibitemOpen
  \bibfield  {author} {\bibinfo {author} {\bibfnamefont {P.}~\bibnamefont
  {Cassak}}, \bibinfo {author} {\bibfnamefont {Y.-H.}\ \bibnamefont {Liu}}, \
  and\ \bibinfo {author} {\bibfnamefont {M.}~\bibnamefont {Shay}},\ }\href@noop
  {} {\bibfield  {journal} {\bibinfo  {journal} {J. Plasma Phys.}\ }\textbf
  {\bibinfo {volume} {83}},\ \bibinfo {pages} {715830501} (\bibinfo {year}
  {2017})}\BibitemShut {NoStop}%
\bibitem [{\citenamefont {Sharma~Pyakurel}\ \emph {et~al.}(2019)\citenamefont
  {Sharma~Pyakurel}, \citenamefont {Shay}, \citenamefont {Phan}, \citenamefont
  {Matthaeus}, \citenamefont {Drake}, \citenamefont {TenBarge}, \citenamefont
  {Haggerty}, \citenamefont {Klein}, \citenamefont {Cassak}, \citenamefont
  {Parashar}, \citenamefont {Swisdak},\ and\ \citenamefont
  {Chasapis}}]{pyakurel2019}%
  \BibitemOpen
  \bibfield  {author} {\bibinfo {author} {\bibfnamefont {P.}~\bibnamefont
  {Sharma~Pyakurel}}, \bibinfo {author} {\bibfnamefont {M.~A.}\ \bibnamefont
  {Shay}}, \bibinfo {author} {\bibfnamefont {T.~D.}\ \bibnamefont {Phan}},
  \bibinfo {author} {\bibfnamefont {W.~H.}\ \bibnamefont {Matthaeus}}, \bibinfo
  {author} {\bibfnamefont {J.~F.}\ \bibnamefont {Drake}}, \bibinfo {author}
  {\bibfnamefont {J.~M.}\ \bibnamefont {TenBarge}}, \bibinfo {author}
  {\bibfnamefont {C.~C.}\ \bibnamefont {Haggerty}}, \bibinfo {author}
  {\bibfnamefont {K.~G.}\ \bibnamefont {Klein}}, \bibinfo {author}
  {\bibfnamefont {P.~A.}\ \bibnamefont {Cassak}}, \bibinfo {author}
  {\bibfnamefont {T.~N.}\ \bibnamefont {Parashar}}, \bibinfo {author}
  {\bibfnamefont {M.}~\bibnamefont {Swisdak}}, \ and\ \bibinfo {author}
  {\bibfnamefont {A.}~\bibnamefont {Chasapis}},\ }\href {\doibase
  10.1063/1.5090403} {\bibfield  {journal} {\bibinfo  {journal} {Physics of
  Plasmas}\ }\textbf {\bibinfo {volume} {26}},\ \bibinfo {pages} {082307}
  (\bibinfo {year} {2019})},\ \Eprint
  {http://arxiv.org/abs/https://doi.org/10.1063/1.5090403}
  {https://doi.org/10.1063/1.5090403} \BibitemShut {NoStop}%
\bibitem [{\citenamefont {Ji}\ and\ \citenamefont {Daughton}(2011)}]{ji11}%
  \BibitemOpen
  \bibfield  {author} {\bibinfo {author} {\bibfnamefont {H.}~\bibnamefont
  {Ji}}\ and\ \bibinfo {author} {\bibfnamefont {W.}~\bibnamefont {Daughton}},\
  }\href@noop {} {\bibfield  {journal} {\bibinfo  {journal} {Phys. Plasmas}\
  }\textbf {\bibinfo {volume} {18}},\ \bibinfo {eid} {111207} (\bibinfo {year}
  {2011})}\BibitemShut {NoStop}%
\bibitem [{\citenamefont {{Vilmer}}(2012)}]{vilmer12}%
  \BibitemOpen
  \bibfield  {author} {\bibinfo {author} {\bibfnamefont {N.}~\bibnamefont
  {{Vilmer}}},\ }\href@noop {} {\bibfield  {journal} {\bibinfo  {journal}
  {Philosophical Transactions of the Royal Society of London Series A}\
  }\textbf {\bibinfo {volume} {370}},\ \bibinfo {pages} {3241} (\bibinfo {year}
  {2012})}\BibitemShut {NoStop}%
\bibitem [{\citenamefont {{Raymond}}\ \emph {et~al.}(2012)\citenamefont
  {{Raymond}}, \citenamefont {{Krucker}}, \citenamefont {{Lin}},\ and\
  \citenamefont {{Petrosian}}}]{raymond12}%
  \BibitemOpen
  \bibfield  {author} {\bibinfo {author} {\bibfnamefont {J.~C.}\ \bibnamefont
  {{Raymond}}}, \bibinfo {author} {\bibfnamefont {S.}~\bibnamefont
  {{Krucker}}}, \bibinfo {author} {\bibfnamefont {R.~P.}\ \bibnamefont
  {{Lin}}}, \ and\ \bibinfo {author} {\bibfnamefont {V.}~\bibnamefont
  {{Petrosian}}},\ }\href@noop {} {\bibfield  {journal} {\bibinfo  {journal}
  {Space Science Rev.}\ }\textbf {\bibinfo {volume} {173}},\ \bibinfo {pages}
  {197} (\bibinfo {year} {2012})}\BibitemShut {NoStop}%
\bibitem [{\citenamefont {Goodman}\ and\ \citenamefont
  {Uzdensky}(2008)}]{goodman08}%
  \BibitemOpen
  \bibfield  {author} {\bibinfo {author} {\bibfnamefont {J.}~\bibnamefont
  {Goodman}}\ and\ \bibinfo {author} {\bibfnamefont {D.}~\bibnamefont
  {Uzdensky}},\ }\href@noop {} {\bibfield  {journal} {\bibinfo  {journal}
  {Astrophys. J.}\ }\textbf {\bibinfo {volume} {688}},\ \bibinfo {pages} {555}
  (\bibinfo {year} {2008})}\BibitemShut {NoStop}%
\bibitem [{\citenamefont {{Cangemi}}\ \emph {et~al.}(2021)\citenamefont
  {{Cangemi}}, \citenamefont {{Rodriguez}}, \citenamefont {{Grinberg}},
  \citenamefont {{Belmont}}, \citenamefont {{Laurent}},\ and\ \citenamefont
  {{Wilms}}}]{cangemi21}%
  \BibitemOpen
  \bibfield  {author} {\bibinfo {author} {\bibfnamefont {F.}~\bibnamefont
  {{Cangemi}}}, \bibinfo {author} {\bibfnamefont {J.}~\bibnamefont
  {{Rodriguez}}}, \bibinfo {author} {\bibfnamefont {V.}~\bibnamefont
  {{Grinberg}}}, \bibinfo {author} {\bibfnamefont {R.}~\bibnamefont
  {{Belmont}}}, \bibinfo {author} {\bibfnamefont {P.}~\bibnamefont
  {{Laurent}}}, \ and\ \bibinfo {author} {\bibfnamefont {J.}~\bibnamefont
  {{Wilms}}},\ }\href@noop {} {\bibfield  {journal} {\bibinfo  {journal}
  {Astro. Astrophys.}\ }\textbf {\bibinfo {volume} {645}},\ \bibinfo {eid}
  {A60} (\bibinfo {year} {2021})}\BibitemShut {NoStop}%
\bibitem [{\citenamefont {{Sari}}\ and\ \citenamefont
  {{Piran}}(1999)}]{sari99}%
  \BibitemOpen
  \bibfield  {author} {\bibinfo {author} {\bibfnamefont {R.}~\bibnamefont
  {{Sari}}}\ and\ \bibinfo {author} {\bibfnamefont {T.}~\bibnamefont
  {{Piran}}},\ }\href@noop {} {\bibfield  {journal} {\bibinfo  {journal}
  {Astrophys. J.}\ }\textbf {\bibinfo {volume} {520}},\ \bibinfo {pages} {641}
  (\bibinfo {year} {1999})}\BibitemShut {NoStop}%
\bibitem [{\citenamefont {Beloborodov}(2017)}]{beloborodov17}%
  \BibitemOpen
  \bibfield  {author} {\bibinfo {author} {\bibfnamefont {A.~M.}\ \bibnamefont
  {Beloborodov}},\ }\href {\doibase 10.3847/2041-8213/aa78f3} {\bibfield
  {journal} {\bibinfo  {journal} {Astrophys. J.}\ }\textbf {\bibinfo {volume}
  {843}},\ \bibinfo {pages} {L26} (\bibinfo {year} {2017})}\BibitemShut
  {NoStop}%
\bibitem [{\citenamefont {{Torricelli-Ciamponi}}, \citenamefont {{Pietrini}},\
  and\ \citenamefont {{Orr}}(2005)}]{torricelli05}%
  \BibitemOpen
  \bibfield  {author} {\bibinfo {author} {\bibfnamefont {G.}~\bibnamefont
  {{Torricelli-Ciamponi}}}, \bibinfo {author} {\bibfnamefont {P.}~\bibnamefont
  {{Pietrini}}}, \ and\ \bibinfo {author} {\bibfnamefont {A.}~\bibnamefont
  {{Orr}}},\ }\href@noop {} {\bibfield  {journal} {\bibinfo  {journal} {Astro.
  Astrophys.}\ }\textbf {\bibinfo {volume} {438}},\ \bibinfo {pages} {55}
  (\bibinfo {year} {2005})}\BibitemShut {NoStop}%
\bibitem [{\citenamefont {{Massaro}}\ and\ \citenamefont
  {{Ajello}}(2011)}]{massaro11}%
  \BibitemOpen
  \bibfield  {author} {\bibinfo {author} {\bibfnamefont {F.}~\bibnamefont
  {{Massaro}}}\ and\ \bibinfo {author} {\bibfnamefont {M.}~\bibnamefont
  {{Ajello}}},\ }\href@noop {} {\bibfield  {journal} {\bibinfo  {journal}
  {Astrophys. J. Lett.}\ }\textbf {\bibinfo {volume} {729}},\ \bibinfo {eid}
  {L12} (\bibinfo {year} {2011})}\BibitemShut {NoStop}%
\bibitem [{\citenamefont {{Kataoka}}\ \emph {et~al.}(2003)\citenamefont
  {{Kataoka}}, \citenamefont {{Edwards}}, \citenamefont {{Georganopoulos}},
  \citenamefont {{Takahara}},\ and\ \citenamefont {{Wagner}}}]{kataoka03}%
  \BibitemOpen
  \bibfield  {author} {\bibinfo {author} {\bibfnamefont {J.}~\bibnamefont
  {{Kataoka}}}, \bibinfo {author} {\bibfnamefont {P.}~\bibnamefont
  {{Edwards}}}, \bibinfo {author} {\bibfnamefont {M.}~\bibnamefont
  {{Georganopoulos}}}, \bibinfo {author} {\bibfnamefont {F.}~\bibnamefont
  {{Takahara}}}, \ and\ \bibinfo {author} {\bibfnamefont {S.}~\bibnamefont
  {{Wagner}}},\ }\href@noop {} {\bibfield  {journal} {\bibinfo  {journal}
  {Astro. Astrophys.}\ }\textbf {\bibinfo {volume} {399}},\ \bibinfo {pages}
  {91} (\bibinfo {year} {2003})}\BibitemShut {NoStop}%
\bibitem [{\citenamefont {Hillas}(1984)}]{hillas_1984}%
  \BibitemOpen
  \bibfield  {author} {\bibinfo {author} {\bibfnamefont {A.~M.}\ \bibnamefont
  {Hillas}},\ }\href {\doibase 10.1146/annurev.aa.22.090184.002233} {\bibfield
  {journal} {\bibinfo  {journal} {Annual Review of Astronomy and Astrophysics}\
  }\textbf {\bibinfo {volume} {22}},\ \bibinfo {pages} {425} (\bibinfo {year}
  {1984})},\ \Eprint
  {http://arxiv.org/abs/https://doi.org/10.1146/annurev.aa.22.090184.002233}
  {https://doi.org/10.1146/annurev.aa.22.090184.002233} \BibitemShut {NoStop}%
\bibitem [{\citenamefont {Guo}\ \emph {et~al.}(2015)\citenamefont {Guo},
  \citenamefont {Liu}, \citenamefont {Daughton},\ and\ \citenamefont
  {Li}}]{guo15}%
  \BibitemOpen
  \bibfield  {author} {\bibinfo {author} {\bibfnamefont {F.}~\bibnamefont
  {Guo}}, \bibinfo {author} {\bibfnamefont {Y.-H.}\ \bibnamefont {Liu}},
  \bibinfo {author} {\bibfnamefont {W.}~\bibnamefont {Daughton}}, \ and\
  \bibinfo {author} {\bibfnamefont {H.}~\bibnamefont {Li}},\ }\href@noop {}
  {\bibfield  {journal} {\bibinfo  {journal} {Astrophys. J.}\ }\textbf
  {\bibinfo {volume} {806}},\ \bibinfo {pages} {1} (\bibinfo {year}
  {2015})}\BibitemShut {NoStop}%
\bibitem [{\citenamefont {Habara}\ \emph {et~al.}(2019)\citenamefont {Habara},
  \citenamefont {Iwawaki}, \citenamefont {Gong}, \citenamefont {Wei},
  \citenamefont {Ivancic}, \citenamefont {Theobald}, \citenamefont {Krauland},
  \citenamefont {Zhang}, \citenamefont {Fiksel},\ and\ \citenamefont
  {Tanaka}}]{habara_2019_ouesm}%
  \BibitemOpen
  \bibfield  {author} {\bibinfo {author} {\bibfnamefont {H.}~\bibnamefont
  {Habara}}, \bibinfo {author} {\bibfnamefont {T.}~\bibnamefont {Iwawaki}},
  \bibinfo {author} {\bibfnamefont {T.}~\bibnamefont {Gong}}, \bibinfo {author}
  {\bibfnamefont {M.~S.}\ \bibnamefont {Wei}}, \bibinfo {author} {\bibfnamefont
  {S.~T.}\ \bibnamefont {Ivancic}}, \bibinfo {author} {\bibfnamefont
  {W.}~\bibnamefont {Theobald}}, \bibinfo {author} {\bibfnamefont {C.~M.}\
  \bibnamefont {Krauland}}, \bibinfo {author} {\bibfnamefont {S.}~\bibnamefont
  {Zhang}}, \bibinfo {author} {\bibfnamefont {G.}~\bibnamefont {Fiksel}}, \
  and\ \bibinfo {author} {\bibfnamefont {K.~A.}\ \bibnamefont {Tanaka}},\
  }\href {\doibase 10.1063/1.5088529} {\bibfield  {journal} {\bibinfo
  {journal} {Review of Scientific Instruments}\ }\textbf {\bibinfo {volume}
  {90}},\ \bibinfo {pages} {063501} (\bibinfo {year} {2019})},\ \Eprint
  {http://arxiv.org/abs/https://doi.org/10.1063/1.5088529}
  {https://doi.org/10.1063/1.5088529} \BibitemShut {NoStop}%
\bibitem [{\citenamefont {Bonnet}\ \emph {et~al.}(2013)\citenamefont {Bonnet},
  \citenamefont {Comet}, \citenamefont {Denis-Petit}, \citenamefont {Gobet},
  \citenamefont {Hannachi}, \citenamefont {Tarisien}, \citenamefont
  {Versteegen},\ and\ \citenamefont {Aléonard}}]{bonnet_2013_ip}%
  \BibitemOpen
  \bibfield  {author} {\bibinfo {author} {\bibfnamefont {T.}~\bibnamefont
  {Bonnet}}, \bibinfo {author} {\bibfnamefont {M.}~\bibnamefont {Comet}},
  \bibinfo {author} {\bibfnamefont {D.}~\bibnamefont {Denis-Petit}}, \bibinfo
  {author} {\bibfnamefont {F.}~\bibnamefont {Gobet}}, \bibinfo {author}
  {\bibfnamefont {F.}~\bibnamefont {Hannachi}}, \bibinfo {author}
  {\bibfnamefont {M.}~\bibnamefont {Tarisien}}, \bibinfo {author}
  {\bibfnamefont {M.}~\bibnamefont {Versteegen}}, \ and\ \bibinfo {author}
  {\bibfnamefont {M.~M.}\ \bibnamefont {Aléonard}},\ }\href {\doibase
  10.1063/1.4826084} {\bibfield  {journal} {\bibinfo  {journal} {Review of
  Scientific Instruments}\ }\textbf {\bibinfo {volume} {84}},\ \bibinfo {pages}
  {103510} (\bibinfo {year} {2013})},\ \Eprint
  {http://arxiv.org/abs/https://doi.org/10.1063/1.4826084}
  {https://doi.org/10.1063/1.4826084} \BibitemShut {NoStop}%
\bibitem [{\citenamefont {Li}\ \emph {et~al.}(2018)\citenamefont {Li},
  \citenamefont {Guo}, \citenamefont {Li},\ and\ \citenamefont
  {Birn}}]{Li_2018}%
  \BibitemOpen
  \bibfield  {author} {\bibinfo {author} {\bibfnamefont {X.}~\bibnamefont
  {Li}}, \bibinfo {author} {\bibfnamefont {F.}~\bibnamefont {Guo}}, \bibinfo
  {author} {\bibfnamefont {H.}~\bibnamefont {Li}}, \ and\ \bibinfo {author}
  {\bibfnamefont {J.}~\bibnamefont {Birn}},\ }\href {\doibase
  10.3847/1538-4357/aaacd5} {\bibfield  {journal} {\bibinfo  {journal} {The
  Astrophysical Journal}\ }\textbf {\bibinfo {volume} {855}},\ \bibinfo {pages}
  {80} (\bibinfo {year} {2018})}\BibitemShut {NoStop}%
\bibitem [{\citenamefont {Daughton}, \citenamefont {Scudder},\ and\
  \citenamefont {Karimabadi}(2006)}]{daughton_openbc_2006}%
  \BibitemOpen
  \bibfield  {author} {\bibinfo {author} {\bibfnamefont {W.}~\bibnamefont
  {Daughton}}, \bibinfo {author} {\bibfnamefont {J.}~\bibnamefont {Scudder}}, \
  and\ \bibinfo {author} {\bibfnamefont {H.}~\bibnamefont {Karimabadi}},\
  }\href {\doibase 10.1063/1.2218817} {\bibfield  {journal} {\bibinfo
  {journal} {Physics of Plasmas}\ }\textbf {\bibinfo {volume} {13}},\ \bibinfo
  {pages} {072101} (\bibinfo {year} {2006})},\ \Eprint
  {http://arxiv.org/abs/https://doi.org/10.1063/1.2218817}
  {https://doi.org/10.1063/1.2218817} \BibitemShut {NoStop}%
\end{thebibliography}%
\clearpage
\newpage
\section*{Supplemental Material}

\subsection*{Study on LPI-generated electron deflections by coil magnetic fields}
The angular dependence of accelerated electrons measured by OU-ESM was interpreted as evidence of electrons accelerated by the reconnection electric field. However, this interpretation does not take into account the possibility of LPI-generated electrons being deflected by coil magnetic fields preferentially in certain angles, resulting in the measured angular distribution in the electron energy spectra. To explore this possibility, numerical raytracing is performed using positrons advanced through the vacuum coil magnetic field.

Instead of advancing electrons from the laser spot to the electron spectrometer, positrons are initialized at the spectrometer, and advanced toward the capacitor-coil target. Due to charge-parity-time (CPT) symmetry, a positron advancing ``backward'' in time through electromagnetic fields is equivalent to an electron advancing ``forward'' in time through the same fields. These two simulations are thus functionally similar, but the former allows for significantly faster computation times due to more limited angular spread around the collimator at the OU-ESM. Positron raytracing is performed for all $5$ OU-ESM channels. Positrons with kinetic energy $50~\textrm{keV}$ are initialized at the entrance of each channel, with an angular spread of $1^{\circ}$ in both polar and azimuthal angles, providing a field of view that adequately encompasses the entire capacitor coil target. These positrons are then advanced through the vacuum magnetic fields calculated from the coil current geometry via a 4th-order Runge-Kutta algorithm. The majority of these positrons leave the field region, but the few that impact the capacitor coil plates are recorded. In addition to scanning across the OU-ESM channel angles, the raytracing is scanned across various coil currents, from $0~\textrm{kA}$, up to the maximum coil current of $57~\textrm{kA}$, and 3 different coil configurations: double coil (reconnection), left coil only (control), and right coil only (control). For each combination of OU-ESM channel, coil current, and coil configuration, $5\times10^{5}$ positrons are used.

\begin{figure*}[htb!]
  \begin{center}
    \includegraphics[width=\textwidth]{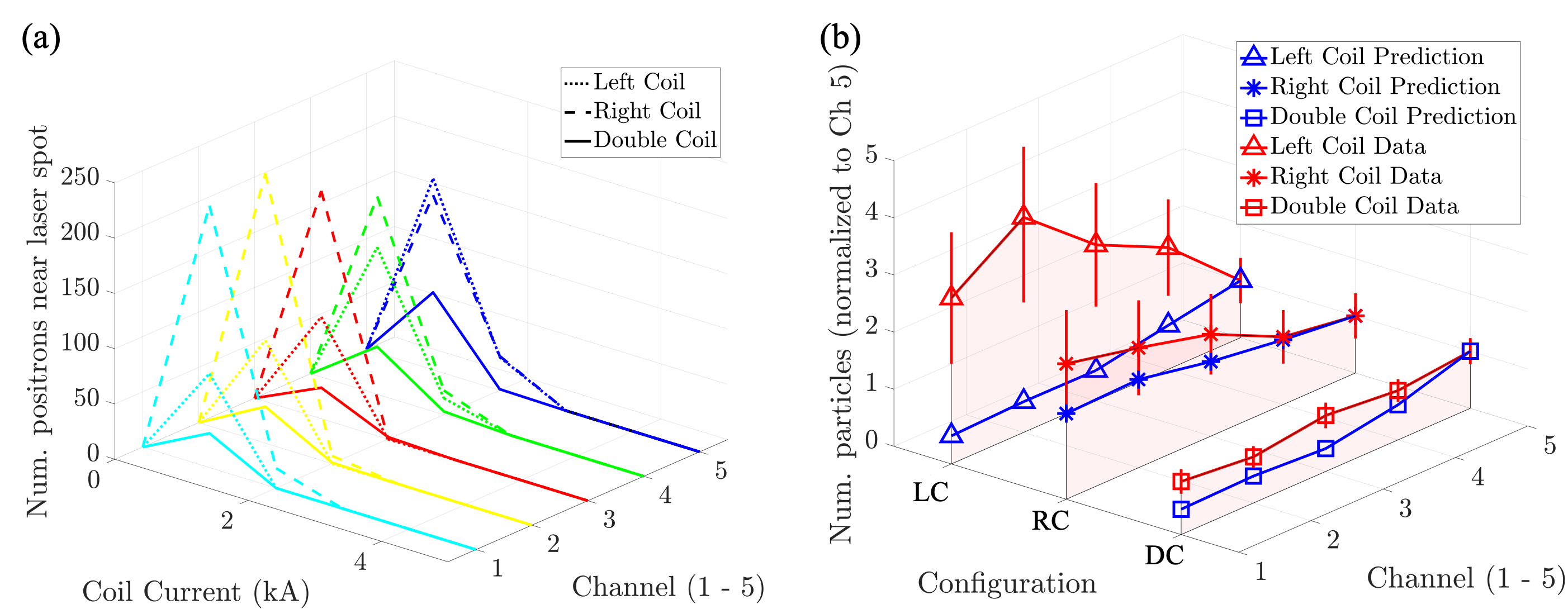}
  \end{center}
  \caption{\textbf{Positron raytracing with vacuum coil magnetic fields without reconnection demonstrate that LPI electrons, deflected by left-coil, right-coil, and double-coil magnetic fields are unable to explain the observed experimental angular trends.} \textbf{a,} The number of positrons near the laser spot are plotted as a function of OU-ESM channel, coil configuration, and coil current, via raytracing simulations. In all coil configurations, positrons are deflected near the laser spot only for small finite coil currents $0<I\lesssim3~\textrm{kA}$, as larger coil currents deflect the positrons below the bottom of the back plate. Further, across all channels, left and right coil configurations consistently exhibit larger positron impacts than the double coil configuration. This exercise implies that LPI-generated electrons are unlikely to be preferentially accelerated to the OU-ESM by double coil magnetic fields, as compared to single coil magnetic fields. \textbf{b,} Inter-channel trends for left coil, right coil, and double coil configurations are drawn from LPI-only predictions (blue) and compared to experimental data. Both predicted and experimental values are normalized to the Channel 5 signal, and the Channel 5 data points are therefore unity, by definition. \textit{Predicted} trends are taken from the coil current $I=1~\textrm{kA}$ case. \textit{Experimental} trends are calculated by taking the ratio of the OU-ESM signal from different channels. Left coil and right coil data are calculated from the respective single shots, while the double coil data are calculated as an average of all three reconnection shots. While double coil and right coil data are somewhat consistent with the raytracing LPI-only predictions, the left coil experimental ratios vary significantly from these predictions.}
  \label{fig:6}
\end{figure*}

In all coil configurations, positrons are only deflected toward the laser spot (defined as a $50~\mu\textrm{m}$-radius circle centered on the back plate) for small finite coil currents $0<I\lesssim3~\textrm{kA}$. For larger coil currents ($I\gtrsim5~\textrm{kA}$), the positrons are deflected completely away from the back plate, with no deposited positrons recorded. For $I\lesssim5~\textrm{kA}$, the deposition pattern of positrons is seen to move ``downward'' with increasing coil current. This is explained by the $\vec{v}\times\vec{B}$ force from the ``horizontal'' magnetic field (left to right in Fig.~\ref{fig:1}b).

An interesting trend appears when comparing the double coil reconnection configurations with either right or left coil control configurations. Across all channels, the right and left coil configurations (represented by triangle and asterisk markers, respectively) result in more positrons deflected near the laser spot than the double coil configuration (represented by square markers). Reverting to the electron frame, this implies that LPI-generated electrons with energies of $50~\textrm{keV}$ are less likely to be deflected toward the electron spectrometer by double coil configurations, compared to single coil configurations. The possibility of the double coil case acting as an ``energy selector'' causing the spectral bumps is therefore small. Furthermore, we have shown that $50~\textrm{keV}$ electrons are deflected toward the OU-ESM for a very limited range of coil currents, when compared to the maximum coil current of $57~\textrm{kA}$. As OU-ESM is a time-integrated diagnostic, even if some sort of energy selection mechanism were to exist, the effect on the entire spectrum would be small, as the coil current spends comparatively little time in the $0<I\lesssim3~\textrm{kA}$ range.

The model is not perfect, primarily due to the assumption of vacuum magnetic fields and the ignoring of plasma effects. The relative importance of plasma effects in the exercise can be illustrated by the no-coil experimental spectra shown in Fig.~\ref{fig:2}f. Without plasma effects, no LPI electrons are expected at the OU-ESM, as the field of view of the instrument do not include the laser spot, from which LPI electrons originate. The low signal level of the measured no-coil spectra therefore illustrates the relative insignificance of plasma effects on measured and simulated electron spectra. Despite the assumptions, the exercise overall does not support an energy selection mechanism, whereby LPI-generated electrons of energy $\sim50~\textrm{keV}$ are deflected by coil magnetic fields alone and are responsible for the spectral bump.

Inter-channel trends can also be predicted by the raytracing exercise, and compared to experimental values. For instance, Fig.~\ref{fig:6}a shows an increase in double coil signal with increasing channel. These LPI-only predictions can be established for all three configurations for coil current $I=1~\textrm{kA}$. Similarly, experimental signal levels for electron energy $E=50~\textrm{keV}$ can be compared across channels for these three configurations. The results are summarized in Fig.~\ref{fig:6}b. The double coil experimental ratios roughly agree with the LPI-only predicted values, hovering close to a ratio of unity, when compared to Channel 5. The right coil configuration shows agreement between LPI-predicted and experimental trends, helped by large error bars in the data. The left coil data, however, demonstrates a large deviation from raytracing predictions. This inconsistency shows that the experimental results \textit{cannot} be adequately explained with vacuum magnetic fields alone deflecting electrons toward different detector channels.

\end{document}